\documentclass[aps,preprint,showpacs]{revtex4}
\usepackage{amsmath}
\usepackage{pstricks}
\usepackage{color}
\usepackage{graphicx}
\newcommand{\be}{\begin{equation}}
\newcommand{\ee}{\end{equation}}
\newcommand{\bea}{\begin{eqnarray}}
\newcommand{\eea}{\end{eqnarray}} 
\newcommand{\ba}{\begin{array}}
\newcommand{\ea}{\end{array}}

\newcommand{\bb}{\bibitem}

\begin{document}
%\draft
%\tightenlines

\title{\bf Unconventional minimal subtraction and Bogoliubov-Parasyuk-Hepp-Zimmermann:  massive scalar theory 
and critical exponents}
\author{Paulo R. S. Carvalho\footnote{e-mail:prscarvalho@ufpi.edu.br}} 
\affiliation{{\it Departamento de F\'\i sica, Universidade Federal do Piau\'\i, Campus Ministro Petr\^onio Portela, 64049-500, Teresina, PI Brazil}}  
\author{Marcelo M. Leite\footnote{e-mail:mleite@df.ufpe.br}}
\affiliation{{\it Laborat\'orio de F\'\i sica Te\'orica e Computacional, Departamento de F\'\i sica,\\ Universidade Federal de Pernambuco,\\
50670-901, Recife, PE, Brazil}}

%\end{center}
\vspace{0.2cm}
\begin{abstract}
{\it We introduce a simpler although unconventional minimal subtraction renormalization procedure in the case of a massive scalar $\lambda \phi^{4}$ theory in Euclidean space using dimensional regularization. We show that this method is very similar to its counterpart in massless field theory. In particular, the choice of using the bare mass at higher perturbative order instead of employing its tree-level counterpart eliminates all tadpole insertions at that order. As an application, we compute diagrammatically the critical exponents $\eta$ and $\nu$ at least up to two loops. We perform an explicit comparison with the Bogoliubov-Parasyuk-Hepp-Zimmermann ($BPHZ$) method at the same loop order, show that the proposed method requires fewer diagrams and establish a connection between 
the two approaches.}

\end{abstract}
\vspace{1cm}

\maketitle

\newpage
\section{Introduction}

\par Methods in field theory are ubiquitous in several areas of theoretical 
physics. The advent of renormalization ideas \cite{GL} and the 
renormalization-group arguments \cite{Wilson} set the ground to various 
schemes in which one can extract finite (from otherwise meaningless infinite) 
results using perturbation theory. An important application is the computation 
of critical exponents from diagrammatic methods in a $\lambda\phi^{4}$ theory 
which describes the universality classes of ordinary systems undergoing a 
phase transition \cite{WK}. During that early stage, the regularization method 
invented to handle properly the infinities due to ultraviolet divergences 
appearing in the computation of Feynman graphs utilized a momentum cutoff 
\cite{IIM}. In addition, the utilization of dimensional regularization 
\cite{tHV,BLZ,LA} together with minimal subtraction of dimensional poles led 
the subject of perturbative computation of exponents using massive fields to 
unprecedented precision within the $\epsilon$-expansion \cite{Kleinert}.
\par Minimal subtraction is much simpler when formulated in terms of massless 
fields. Consider only the multiplicatively renormalizable one-particle 
irreducible ($1PI$) vertex functions. The massless integrals are easier to 
evaluate since all polynomial dependence of a given diagram in the mass 
vanishes. Consequently, the minimal subtraction approach can be formulated 
without the need of the iterative construction of counterterms \cite{Amit}. In 
case of a massive theory, one has to employ the technique named ``partial $p$'' \cite{tHV} in order to separate the dependence of polynomials in the mass from 
the contribution of polynomials in the external momenta (beside the 
contributions of logarithmical integrals combining both). A standard procedure 
is to employ the $BPHZ$ renormalization method, which is the statement of 
renormalizability on the level of the Lagrangian\cite{BP,BS,H,Z1}. The bare Lagrangian density 
includes the counterterms to be constructed perturbatively with 
appropriate vertices such that the theory is automatically 
renormalized. The counterterms determine three normalization constants, 
corresponding to the renormalization of the field, the mass and coupling 
constant, respectively. Within this scheme we have to compute a large number of diagrams. We can then ask 
ourselves whether we can find out a much simpler minimal subtraction version 
with a reduced number of diagrams in a certain higher order in the number of loops and verify its consistency 
with the rigorous but lenghtier $BPHZ$ technique through a simple application.
\par In this work we propose a new method of minimal subtraction of 
dimensional poles in a massive $\lambda \phi^{4}$. We restrict our discussion 
of the various vertex parts involved up to two-loop level, except by the 
two-point function, which is going to be determined up to three-loop level. 
In the bare propagators, 
the tree-level bare mass is replaced by the three-loop bare mass. At two-loop 
level, the vertex parts which are required to renormalize the theory 
multiplicatively at this perturbative order work pretty much in the same 
manner as in the massless theory with the elimination of all tadpole graphs 
present in the diagrammatic expansion at the same perturbative order. We just 
need the renormalization functions of the field, the composite field and the 
expansion of the tree-level dimensionless bare coupling constant in terms of 
the renormalized dimensionless coupling. The nontrivial feature of this 
unconventional approach is that the two-point vertex part at three-loop order 
requires an extra renormalization, since the choice of the new bare mass 
parameter produces a residual single pole in that vertex function. 
\par As an application, we compute the anomalous dimension of the field 
$\eta$ at three-loop order as well as the correlation length exponent $\nu$ 
at two-loop order by diagrammatic means. We then perform a detailed comparison 
with the traditional $BPHZ$ method of minimal subtraction: we evaluated 
explicitly all the required diagrams, the fixed point and the corresponding 
Wilson functions at the fixed point. In spite of quite different intermediate 
results, we show that our method is much simpler 
since we only need to calculate a reduced number of diagrams. Comparing those 
results we are led to a dictionary between the two minimal subtraction 
renormalization schemes for this massive scalar field theory.     
\par Section II presents all primitive divergent vertex parts required for 
the multiplicative renormalization along with their diagrammatic loop 
expansion. The minimal set of integrals (and their solutions as poles in 
$\epsilon$) displayed in this section is outlined in 
Appendix A. Section III deals with the explicit renormalization of the vertex 
parts and how the extra subtraction in the two-point function at three-loops 
can be performed without changing the various normalization constants. We 
compute the Wilson functions, the fixed point and give a brief 
description of the computation of the critical exponents. 
\par The $BPHZ$ method is reviewed in Section IV. We calculate the 
critical exponents using this technique. Owing to simplicity, in Appendix B we evaluate one 
three-loop integral of the two-point function along with its counterterm 
in order to prove that the singular part of that combination of diagrams does 
not depend on the external momentum. 
\par In Section V we discuss our proposal and include possible future 
applications in the concluding remarks. 

\section{Unconventional minimal subtraction for the massive theory}

\par Originally, the method which requires a minimal number of diagrams with 
the elimination of all tadpoles along the way using dimensional regularization 
was already discussed in Amit and Martin-Mayor's book \cite{Amit} in the 
framework of renormalized massless fields. Our goal here is to adapt that 
technique to the massive renormalized theory. We shall employ that 
notation throughout this work. 
\par The bare Lagrangian with $O(N)$-symmetry we are going to consider is 
given by 
\begin{equation}\label{1}
\mathcal{L} = \frac{1}{2}
|\bigtriangledown \phi_{0}|^{2} + \frac{1}{2} \mu_{0}^{2}\phi_0^{2} + \frac{1}{4!}\lambda(\phi_0^{2})^{2} ,
\end{equation}
where $\phi_{0}$, $\mu_{0}$ and $\lambda$ are the bare order parameter, 
mass ($\mu_{0}^{2}= t_{0}$ is the bare reduced temperature in statistical 
mechanics, which is proportional to $\frac{T-T_{C}}{T_{C}}$) and coupling 
constant, respectively. Note that this expression tells us that we are going 
to proceed with our discussion in Euclidean space due to the signature chosen 
for the quadratic term in the derivatives. Our discussion will be founded entirely in momentum space. The primitively (bare) divergent 
one-particle irreducibe ($1PI$) vertex parts which are required to renormalize the theory 
multiplicatively are the two-point vertex part 
$\Gamma^{(2)}(k_{1},k_{2},\mu_{0},\lambda)$, the four-point vertex 
$\Gamma^{(4)}(k_{1},k_{2},k_{3},k_{4},\mu_{0},\lambda)$ and the 
two-point vertex point with composite field insertion 
$\Gamma^{(2,1)}(k_{1},k_{2};Q_{3},\mu_{0},\lambda)$, where $k_{i}$ are the 
external momenta associated to the particular vertex part and $Q_{3}$ is the 
external momentum of the insertion of the $\phi^{2}$ composite operator. The reader 
should be aware that not all external momenta 
are independent.
\par Let us turn now to the multiplicative renormalizability statement in this 
formulation. One can 
define finite renormalized vertex parts out of the (infinite) bare vertex 
parts with arbitrary insertions of composite 
operators $\Gamma^{(N,M)}(k_{i};Q_{j})$ where $i=1,...,N, 
j=1,...,M$ ($(N,M)\neq(0,2)$) using renormalization functions. When the bare vertices are multiplied by the normalization 
functions of the field $Z_{\phi}$ and of the composite field $Z_{\phi^{2}}$ 
(whose divergences manifest themselves as inverse powers of $\epsilon=4-d$), they yield the renormalized vertex parts 
\begin{equation}\label{2}
\Gamma_{R}^{(N,M)}(k_{i}; Q_{j}, g, m)= (Z_{\phi})^{\frac{N}{2}}
(Z_{\phi^{2}})^{M} \Gamma^{(N,M)}(k_{i}; Q_{j}),
\end{equation} 
which turn out to be finite. Note that in the left hand side of the last 
equation the parameters $m$ and $g$ are the renormalized mass and coupling 
constant, respectively.
\par Now let us discuss what are the bare parameters which enter in the right 
hand side of last equation. In order to do that, let us consider the loop 
expansion of the three vertex parts above mentioned. 
\par Begin with $\Gamma^{(2)}$. Instead of writing the complete expression, 
let us draw the diagrams which correspond to all integrals that are going 
to be required in our subsequent discussion. The simplest diagrams which do 
not depend on the external momenta are the tadpoles and their cousins, which 
are given by the following 
expressions
\begin{eqnarray}
%tadpole
\parbox{10mm}{\includegraphics[scale=1.0]{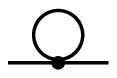}} \quad &=& \frac{N+2}{3}\int \frac{d^{d}q}{q^{2}+\mu_{0}^{2}}, \label{3}\\
%tadpolo duplo
\parbox{10mm}{\includegraphics[scale=1.0]{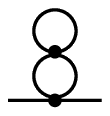}} \quad &=& \left(\frac{N+2}{3} \right)^{2} 
\int \frac{d^{d}q_{1}d^{d}q_{2}}{(q_{1}^{2}+\mu_{0}^{2})^{2}
(q_{2}^{2}+\mu_{0}^{2})}, \label{4}\\
% tadpole triplo
\parbox{9mm}{\includegraphics[scale=1.0]{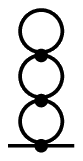}} \quad &=& \left(\frac{N+2}{3} \right)^{3} 
\int \frac{d^{d}q_{1}d^{d}q_{2}d^{d}q_{3}}{(q_{1}^{2}+\mu_{0}^{2})^{2}
(q_{2}^{2}+\mu_{0}^{2})^{2}(q_{3}^{2}+\mu_{0}^{2})}, \label{5}\\
% mickey mouse
\parbox{12mm}{\includegraphics[scale=1.0]{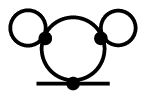}}   \quad &=& \left(\frac{N+2}{3} \right)^{3} 
\int \frac{d^{d}q_{1}d^{d}q_{2}d^{d}q_{3}}{(q_{1}^{2}+\mu_{0}^{2})^{3}
(q_{2}^{2}+\mu_{0}^{2})^{2}(q_{3}^{2}+\mu_{0}^{2})} , \label{6}\\
% bola de basquete
\parbox{10mm}{\includegraphics[scale=1.0]{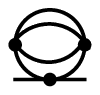}}  \quad &=& \left(\frac{N+2}{3} \right)^{2} 
\int \frac{d^{d}q_{1}d^{d}q_{2}d^{d}q_{3}}{(q_{1}^{2}+\mu_{0}^{2})^{2}
(q_{2}^{2}+\mu_{0}^{2})(q_{3}^{2}+\mu_{0}^{2})((q_{1}+q_{2}+q_{3})^{2}+\mu_{0}^{2})}. \label{7}
\end{eqnarray} 
The other diagrams do depend on the external momenta and can be expressed 
in terms of integrals in the form
\begin{eqnarray}
% sunset
\parbox{10mm}{\includegraphics[scale=1.0]{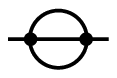}}  \quad &=& \left(\frac{N+2}{3} \right)
\int \frac{d^{d}q_{1}d^{d}q_{2}}{(q_{1}^{2}+\mu_{0}^{2})^{2}
(q_{2}^{2}+\mu_{0}^{2})((q_{1}+q_{2}+k)^{2}+\mu_{0}^{2})},\label{8}\\
% sutian
\parbox{10mm}{\includegraphics[scale=1.0]{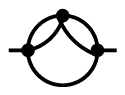}}  \quad &=& \left(\frac{(N+2)(N+8)}{27}\right)
\int \frac{d^{d}q_{1}d^{d}q_{2}d^{d}q_{3}}{(q_{1}^{2}+\mu_{0}^{2})
(q_{2}^{2}+\mu_{0}^{2})(q_{3}^{2}+\mu_{0}^{2})}\nonumber\\
&& \quad \times \quad \frac{1}{((q_{1}+q_{2}+k)^{2}+\mu_{0}^{2})
(q_{1}+q_{3}+k)^{2}+\mu_{0}^{2})}, \label{9}\\
% sunset com tadpole
\parbox{10mm}{\includegraphics[scale=1.0]{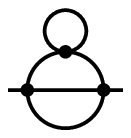}}   \quad &=& \left(\frac{N+2}{3} \right)^{2}
\int \frac{d^{d}q_{1}d^{d}q_{2}d^{d}q_{3}}{(q_{1}^{2}+\mu_{0}^{2})^{2}
(q_{2}^{2}+\mu_{0}^{2})((q_{1}+q_{2}+k)^{2}+\mu_{0}^{2})
(q_{3}^{2}+\mu_{0}^{2})}. \label{10}
\end{eqnarray}
\par Actually we shall not need all diagrams above displayed. In fact, our 
search is to choose a minimal set of graphs to work with.  In a complete 
analogy with the massless framework, we could think of considering only two of 
these diagrams, namely Eqs. (\ref{8}) and (\ref{9}). 
\par Denoting the bare 
propagator $(k^{2} + \mu_{0}^{2})^{-1}$ by a line, we can write a symbolic 
expression of the diagrammatic expansion involving $\Gamma^{(2)}$ at three-loop 
order with respect to our previous graphs. We obtain:  
%Função de 2 Pontos%
\begin{eqnarray}
% propagador livre
\Gamma^{(2)}(k, \mu_{0},\lambda) & = & \quad
\parbox{12mm}{\includegraphics[scale=1.0]{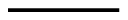}}^{-1} \quad + \quad \frac{\lambda}{2}
% tadpole
\parbox{12mm}{\includegraphics[scale=1.0]{fig1.eps}} \quad - \quad \frac{\lambda^2}{4}
% tadpole duplo
\parbox{12mm}{\includegraphics[scale=1.0]{fig2.eps}} \quad - \quad \frac{\lambda^2}{6}
% sunset
\parbox{12mm}{\includegraphics[scale=1.0]{fig6.eps}}  \quad + \quad  \frac{\lambda^3}{4}
% sutian
\parbox{12mm}{\includegraphics[scale=1.0]{fig7.eps}} \nonumber \\ & & \quad + \quad  \frac{\lambda^3}{4}
% sunset com tadpole
\parbox{12mm}{\includegraphics[scale=1.0]{fig8.eps}} \quad + \quad  \frac{\lambda^3}{12}
% bola de basquete
\parbox{12mm}{\includegraphics[scale=1.0]{fig5.eps}} \quad + \quad  \frac{\lambda^3}{8}
% tadpole triplo
\parbox{12mm}{\includegraphics[scale=1.0]{fig3.eps}} \quad + \quad  \frac{\lambda^3}{8}
% mickey mouse
\parbox{12mm}{\includegraphics[scale=1.0]{fig4.eps}}. \label{11}
\end{eqnarray}
\par Next, define the three-loop bare mass parameter through the expression $\mu=\Gamma^{(2)}(k=0, \mu_{0},\lambda)$, or explicitly
\begin{eqnarray}
\mu^2 & = & \mu_{0}^{2} \quad + \quad \frac{\lambda}{2}
% tadpole
\parbox{12mm}{\includegraphics[scale=1.0]{fig1.eps}} \quad - \quad \frac{\lambda^2}{4}
% tadpole duplo
\parbox{12mm}{\includegraphics[scale=1.0]{fig2.eps}} \quad - \quad \frac{\lambda^2}{6}
% sunset
\parbox{12mm}{\includegraphics[scale=1.0]{fig6.eps}}\bigg|_{k=0}  \quad  + \quad  \frac{\lambda^3}{4}
% sutian
\parbox{12mm}{\includegraphics[scale=1.0]{fig7.eps}}\bigg|_{k=0}  \nonumber \\ & & \quad + \quad
 \frac{\lambda^3}{4}
% sunset com tadpole
\parbox{12mm}{\includegraphics[scale=1.0]{fig8.eps}}\bigg|_{k=0}  \quad + \quad  \frac{\lambda^3}{12}
% bola de basquete
\parbox{12mm}{\includegraphics[scale=1.0]{fig5.eps}} \quad + \quad  \frac{\lambda^3}{8}
% tadpole triplo
\parbox{12mm}{\includegraphics[scale=1.0]{fig3.eps}} \quad + \quad  \frac{\lambda^3}{8}
% mickey mouse
\parbox{12mm}{\includegraphics[scale=1.0]{fig4.eps}}. \label{12}
\end{eqnarray}
\par Performing the inversion to express
$\mu_{0}$ in terms of $\mu$, we get to an expansion for 
$\Gamma^{(2)}(k,\mu_{0},\lambda)$, which depends implicitly on $\mu$. The tadpoles graphs and their cousins will 
vanish into the $\Gamma^{(2)}(k,\mu_{0},\lambda)$ expression obtained after 
the substitution $\mu_{0}(\mu)$. What remain are the diagrams 
Eqs. (\ref{8}), (\ref{9}), (\ref{10}) subtracted from their counterparts 
computed at $k=0$. Expanding $\mu_{0}$ up to first order in 
the coupling constant inside the ``sunset'' diagrams, it follows that 
\begin{eqnarray}
% sunset
\parbox{12mm}{\includegraphics[scale=1.0]{fig6.eps}}\bigg|_{\mu_{0}}  & = & \quad
% sunset
\parbox{12mm}{\includegraphics[scale=1.0]{fig6.eps}}\bigg|_{\mu}  \quad + \quad
\frac{3\lambda}{2}
% sunset com tadpole
\parbox{12mm}{\includegraphics[scale=1.0]{fig8.eps}}\bigg|_{\mu}. \label{13}
\end{eqnarray}
By replacing this expression into $\Gamma^{(2)}(k,\mu_{0},\lambda)$, the 
diagram corresponding to Eq. (\ref{10}) is eliminated. Second, we find out an expression which 
no longer depends on $\mu_{0}$, namely
\begin{eqnarray}
\Gamma^{(2)}(k,\mu_{0},\lambda) & = & k^{2} + \mu^{2} \quad - \quad  \quad
\frac{\lambda^2}{6}\left(
% sunset
\parbox{12mm}{\includegraphics[scale=1.0]{fig6.eps}}\bigg|_{\mu}  \quad - \quad
% sunset
\parbox{12mm}{\includegraphics[scale=1.0]{fig6.eps}}\bigg|_{k=0, \mu}\right)
 \nonumber \\ & & \quad  + \quad  \frac{\lambda^{3}}{4}\left(
% sutian
\parbox{12mm}{\includegraphics[scale=1.0]{fig7.eps}}\bigg|_{\mu}  \quad - \quad
% sutian
\parbox{12mm}{\includegraphics[scale=1.0]{fig7.eps}}\bigg|_{k=0, \mu}\right) . \label{14}
\end{eqnarray}
\par The two-point vertex 
function now depends explicitly on the three-loop bare mass $\mu$. We can write this transmutation as 
$\Gamma(k,\mu_{0},\lambda) \equiv \Gamma(k,\mu,\lambda)$ and no longer have 
to make any reference to the tree-level bare mass. What 
occurs to the other primitively divergent vertex parts when we perform a similar manipulation 
on their graphs?
\par Consider the four-point vertex part. Its diagrams up to two-loop 
order are given by
\begin{eqnarray}
\left[\parbox{10mm}{\includegraphics[scale=1.0]{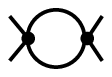}}\right](k) \quad &=& \frac{(N+8)}{9}
\int \frac{d^{d}q}{(q^{2} + \mu_{0}^{2})((q+k)^{2} + \mu_{0}^{2})}, \label{15}\\
\left[\parbox{16mm}{\includegraphics[scale=1.0]{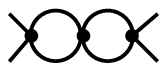}}\right](k) \quad &=& \frac{N^{2}+6N+20}{27}
\int \frac{d^{d}q_{1}d^{d}q_{2}}{(q_{1}^{2} + \mu_{0}^{2})((q_{1}+k)^{2} + \mu_{0}^{2})(q_{2}^{2} + \mu_{0}^{2})}\nonumber\\
&& \qquad \;\times\;\;\frac{1}{((q_{2}+k)^{2} + \mu_{0}^{2})}, \label{16}\\
\left[\parbox{10mm}{\includegraphics[scale=1.0]{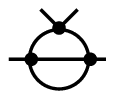}}\right](k_{i}) \quad &=& \left(\frac{5N+22}{27}\right) 
\int\frac{d^{d}q_{1}d^{d}q_{2}}{(q_{1}^{2} + \mu_{0}^{2})((k_{1}+k_{2}-q_{1})^{2} + \mu_{0}^{2})}\nonumber\\
&& \qquad \times \frac{1}{(q_{2}^{2} + \mu_{0}^{2})((k_{3}+q_{1}-q_{2})^{2} + \mu_{0}^{2})}, \label{17}\\
\left[\parbox{10mm}{\includegraphics[scale=1.0]{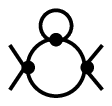}}\right](k) \quad &=& \frac{(N+2)(N+8)}{27} 
\int \frac{d^{d}q_{1}d^{d}q_{2}}{(q_{1}^{2} + \mu_{0}^{2})((q_{1}+k)^{2} + \mu_{0}^{2})^{2}(q_{2}^{2}+ \mu_{0}^{2})}. \label{18}
\end{eqnarray} 
Henceforth, we denote a permutation of external momenta on vertex parts which depend upon them by ``$perm.$''. The expansion of 
the four-point vertex function can be written pictorically as   
%Função de 4 Pontos%
\begin{eqnarray}
&&\Gamma^{(4)}(k_{i},\mu_{0},\lambda) =  \lambda - \frac{\lambda^{2}}{2}
 \left(\left[\parbox{10mm}{\includegraphics[scale=1.0]{fig10.eps}}\right](k_{1}+k_{2}) + 2perms.\right) \nonumber\\
&& + \frac{\lambda^{3}}{4}
         \left(\left[\parbox{16mm}{\includegraphics[scale=1.0]{fig11.eps}}\right](k_{1}+k_{2}) + 2perms.\right) 
+ \frac{\lambda^{3}}{2}
   \left(\left[\parbox{10mm}{\includegraphics[scale=1.0]{fig12.eps}}\right](k_{i}) + 5perms. \right)\nonumber\\
&& + \frac{\lambda^{3}}{2}
  \left(\left[\parbox{10mm}{\includegraphics[scale=1.0]{fig13.eps}}\right](k_{1}+k_{2}) + 2perms. \right). \label{19} 
\end{eqnarray}
With the substitution of the initial bare mass in the 
propapagators of all these graphs as a function of the new bare mass $\mu$, the first diagram will produce itself calculated 
with $\mu$ plus a correction which exactly cancels the last term. After the replacement  $\mu_{0} \rightarrow \mu$  we find
\begin{eqnarray}
&&\Gamma^{(4)}(k_{i},\mu_{0},\lambda) =  \lambda - \frac{\lambda^{2}}{2}
 \left(\left[\parbox{10mm}{\includegraphics[scale=1.0]{fig10.eps}}\right]_{\mu}(k_{1}+k_{2}) + 2perms.\right) \nonumber\\
&& + \frac{\lambda^{3}}{4}
         \left(\left[\parbox{16mm}{\includegraphics[scale=1.0]{fig11.eps}}\right]_{\mu}(k_{1}+k_{2}) + 2perms.\right)
\nonumber\\
&& + \frac{\lambda^{3}}{2}
   \left(\left[\parbox{10mm}{\includegraphics[scale=1.0]{fig12.eps}}\right]_{\mu}(k_{i}) + 5perms. \right). \label{20}
\end{eqnarray}
\par The four-point vertex part ``does not 
remember''the original dependence on $\mu_{0}$ after the elimination of extra 
diagrams and we are left with the minimal set of its diagrams with 
$\mu_{0} \rightarrow \mu$. If we interpret 
$\Gamma^{(4)}(k_{i},\mu,\lambda)(\equiv 
\Gamma^{(4)}(k_{i},\mu,\lambda))$ at two loops as a two-loop truncation of the 
value of $\mu$, it is consistent even though the bare mass 
is defined at three-loop order. 
\par  Finally, the diagrams of the bare vertex $\Gamma^{(2,1)}(k_{1},k_{2};Q_{3},\mu_{0},\lambda)$ up to two-loop order 
are given by:
\begin{eqnarray}
\left[\parbox{13mm}{\includegraphics[scale=1.0]{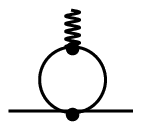}}\right](k) \quad  &=& \frac{N+2}{18}\int \frac{d^{d}q}
{(q^{2}+\mu_{0}^{2})((q+k)^{2} + \mu_{0}^{2})}, \label{21}\\
\left[\parbox{16mm}{\includegraphics[scale=1.0]{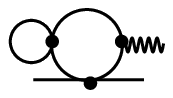}}\right](k) \quad &=& \frac{(N+2)^{2}}{54} 
\int \frac{d^{d}q_{1}d^{d}q_{2}}
{(q_{1}^{2}+\mu_{0}^{2})((q_{1}+k)^{2} + \mu_{0}^{2})(q_{2}^{2}+\mu_{0}^{2})}, 
\label{22}\\
\left[\parbox{11mm}{\includegraphics[scale=1.0]{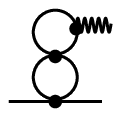}}\right](k) \quad &=& \frac{(N+2)^{2}}{108}\int \frac{d^{d}q_{1}d^{d}q_{2}}
{(q_{1}^{2}+\mu_{0}^{2})((q_{1}+k)^{2} + \mu_{0}^{2})}\nonumber\\
&& \qquad \times \frac{1}{(q_{2}^{2}+\mu_{0}^{2})
((q_{2}+k)^{2} + \mu_{0}^{2})}, \label{23}\\
\left[\parbox{13mm}{\includegraphics[scale=1.0]{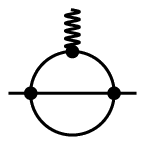}}\right](k_{1},k_{2};Q_{3}) &=& \frac{(N+2)}{36}\int\frac{d^{d}q_{1}d^{d}q_{2}}{(q_{1}^{2} + \mu_{0}^{2})((k_{1}+k_{2}-q_{1})^{2} + \mu_{0}^{2})}\nonumber\\
&& \qquad \times \frac{1}{(q_{2}^{2} + \mu_{0}^{2})((Q_{3}+q_{1}-q_{2})^{2} + \mu_{0}^{2})}. \label{24}
\end{eqnarray}
The diagrammatic expansion for the vertex 
$\Gamma^{(2,1)}(k_{1},k_{2};Q_{3},\mu_{0},\lambda)$ can be written as
\begin{eqnarray}
&& \Gamma^{(2,1)}(k_{1},k_{2};Q_{3},\mu_{0},\lambda) = 1 - 
\lambda \left(\left[\parbox{13mm}{\includegraphics[scale=1.0]{fig14.eps}}\right](k_{1}+k_{2}) + 2 perms.\right)\nonumber\\
&& \quad + \quad \lambda^{2}\left(\left[\parbox{16mm}{\includegraphics[scale=1.0]{fig15.eps}}\right](k_{1}+k_{2}) + 2perms.\right) \quad + \quad \lambda^{2}\left(\left[\parbox{11mm}{\includegraphics[scale=1.0]{fig16.eps}}\right](k_{1}+k_{2}) + 2 perms. \right)\nonumber\\
&& \quad + \quad \lambda^{2}\left(\left[\parbox{13mm}{\includegraphics[scale=1.0]{fig17.eps}}\right](k_{1},k_{2};Q_{3})  + 5 perms.\right).\label{25}
\end{eqnarray}
In those graphs all the propagators are evaluated with $\mu_{0}$ up to now. 
Now expanding $\mu_{0}^{2}(\mu)$ as before, the $O(\lambda)$ first nontrivial 
diagrams originate themselves with $\mu_{0}$ replaced by $\mu$ along with 
$O(\lambda^{2})$ corrections which eliminate precisely  the second diagrams. 
We then get to the following expression:
\begin{eqnarray}
&& \Gamma^{(2,1)}(k_{1},k_{2};Q_{3},\mu_{0},\lambda) = 1 - 
\lambda \left(\left[\parbox{13mm}{\includegraphics[scale=1.0]{fig14.eps}}\right]_{\mu}(k_{1}+k_{2}) + 2perms.\right)\nonumber\\
&& + \lambda^{2}\left(\left[\parbox{11mm}{\includegraphics[scale=1.0]{fig16.eps}}\right]_{\mu}(k_{1}+k_{2}) + 2perms.\right)\;  + \lambda^{2}\left(\left[\parbox{13mm}{\includegraphics[scale=1.0]{fig17.eps}}\right]_{\mu}(k_{1},k_{2};Q_{3})  + 5perms.\right). \label{26}
\end{eqnarray}   
We conclude that the same desirable feature goes on again: the bare vertex 
$\Gamma^{(2,1)}(k_{1},k_{2};Q_{3},\mu_{0},\lambda)$ is insensitive to 
$\mu_{0}$ and if we define 
$\Gamma^{(2,1)}(k_{1},k_{2};Q_{3},\mu,\lambda)\equiv 
\Gamma^{(2,1)}(k_{1},k_{2};Q_{3},\mu_{0},\lambda)$, this vertex is also 
reduced to the minimal number of diagrams with propagator involving only 
the new bare mass parameter $\mu$.
\par Any potentially divergent vertex part which can be renormalized 
multiplicatively includes the three primitively divergent vertex parts just 
discussed (skeleton expansion \cite{ZJ}). Thus, the renormalization of the 
three vertex parts which include only the minimal number of diagrams will be 
sufficient to minimally renormalize the vertex function under consideration. 
Now we can make explicit reference to the argument of the bare vertex 
functions. At first, the would be new multiplicative renormalizability should be a 
statement that by considering a given bare theory 
{\it with bare mass $\mu$} at a certain perturbative order (and a given 
tree-level bare coupling constant), the 
renormalized vertex functions should be finite and satisfy
\begin{equation}\label{27}
\Gamma_{R}^{(N,M)}(k_{i}; Q_{j}, g, m)= (Z_{\phi})^{\frac{N}{2}}
(Z_{\phi^{2}})^{M} \Gamma^{(N,M)}(k_{i}; Q_{j}, \mu,\lambda).
\end{equation} 
\par From now on we are going to determine the normalization functions $Z_{\phi}$ and 
$Z_{\phi^{2}}$ using the results of the computation of the Feynman diagrams 
in Appendix A. Due to our explicit treatment of the bare vertex part 
$\Gamma^{(2,1)}(k_{1},k_{2};Q_{3},\mu,\lambda)$, we define the quantity 
$\bar{Z}_{\phi^{2}}= Z_{\phi^{2}}Z_{\phi}$ which shall be useful in our 
manipulations. 
\par We follow a trend which is standard in the computation of critical 
exponents using this language \cite{Amit}. Nevertheless, we shall see in a moment that this 
new technique does bring new insight in renormalization theory.

\section{Renormalization functions and critical exponents in unconventional 
minimal subtraction}
\par Here we are going to calculate explicitly the critical
exponents. The mainstream of our presentation will be brief, since this material 
is standard in literature \cite{Amit,ZJ}. However, we shall unveil the role of the extra 
subtraction in the renormalized mass and its relationship with certain integrals which will 
come up in our discussion. 
\par The renormalized theory possesses a flow in parameter space generated by the 
renormalized mass $m$: the same bare theory may give origin to many 
renormalized theories with different renormalized masses. The renormalization 
group flow of the coupling constant in parameter 
space is generated by the function 
$\beta(g,m)= m \frac{\partial g}{\partial m}$ . In order to discard 
undesirable dimensionful parameters when $d=4-\epsilon$, define the 
Gell-Mann-Low function $[\beta(g,\mu)]_{GL}= -\epsilon g + \beta(g, \mu)$. 
Hence, even away from the critical dimension we are able to get rid of all 
dimensional couplings defining them in terms of dimensionless couplings 
as $\lambda= \mu^{\epsilon}u_{0}$ and $g=\mu^{\epsilon}u$, where $\mu$ 
is the bare mass at the loop order considered. By using the Gell-Mann-Low function inside the 
Callan-Symanzik equation, the description follows entirely in terms of 
dimensionless  coupling constant. Those definitions imply that 
the object $[\beta(g,\mu)]_{GL} \frac{\partial}{\partial g}= \beta(u) \frac{\partial}
{\partial u}$ has a well defined scaling 
limit \cite{Vladimirov,Naud}. 
\par After collecting these steps together we are 
left with the perturbative computation of the Wilson functions
\begin{subequations}\label{28}
\begin{eqnarray}
\beta(u)&=& -\epsilon \left(\frac{\partial lnu_{0}}{\partial u}\right), \label{28a}\\
\gamma_{\phi}(u) &=& \beta(u)
\left(\frac{\partial ln Z_{\phi}}{\partial u}\right), \label{28b}\\ 
\gamma_{\phi^{2}}(u) &=& - \beta(u) 
\left(\frac{\partial ln Z_{\phi^{2}}}{\partial u}\right), \label{28c}\\
\bar{\gamma}_{\phi^{2}}(u) &=& - \beta(u) 
\left(\frac{\partial ln \bar{Z}_{\phi^{2}}}{\partial u}\right)= 
- \gamma_{\phi}(u) + \gamma_{\phi^{2}}(u). \label{28d}
\end{eqnarray}
\end{subequations} 
\par We first write the 
primitively divergent bare vertex expansion in terms of the minimal 
set of Feynman diagrams previously defined in the form
\begin{subequations}\label{29}
\begin{eqnarray}
&& \Gamma^{(2)}(k, u_{0}, \mu) =
k^{2} + \mu^{2} - B_{2}\mu^{2 \epsilon} u_{0}^{2} + B_{3}\mu^{3\epsilon}u_{0}^{3}, \label{29a}\\
&& \Gamma^{(4)}(k_{i}, u_{0}, \mu) = \mu^{\epsilon} u_{0}
[1- A_{1} \mu^{\epsilon} u_{0}
+ (A_{2}^{(1)} + A_{2}^{(2)})\mu^{2 \epsilon} u_{0}^{2}], \label{29b}\\
&& \Gamma^{(2,1)}(k_{1}, k_{2}; p,u_{0}, \mu) = 1 - C_{1} \mu^{\epsilon} u_{0} 
+ (C_{2}^{(1)} + C_{2}^{(2)}) \mu^{2 \epsilon} u_{0}^{2}. \label{29c}
\end{eqnarray}
\end{subequations}
Since the main modification with respect to standard minimal subtraction 
schemes in the computation of critical exponents is related to the two-point 
vertex part, we consider it as our starting point. It is obvious from Eqs. (\ref{14}) along with Eqs. (\ref{A29}), (\ref{A32}) and 
(\ref{A33}) from Appendix A that 
\begin{equation}\label{30}
B_{2}= \frac{N+2}{18} \mu^{-2\epsilon}\Bigl \{ - \frac{P^{2}}{8\epsilon}\Bigr[1+\frac{1}{4}\epsilon-2\epsilon
\tilde{L}_{3}(P,\mu)\Bigr] - \frac{3\mu^{2}}{4} \tilde{I}(P)\Bigr\}.
\end{equation}
Furthermore, comparing Eq. (\ref{29a}) with the diagrammatic expansion (\ref{14}), we 
obtain the following symbolic result
\begin{equation}\label{31}
B_{3}=  \frac{1}{4}\left(
% sutian
    \parbox{12mm}{\includegraphics[scale=1.0]{fig7.eps}}\bigg|_{\mu}  \quad - \quad
% sutian
    \parbox{12mm}{\includegraphics[scale=1.0]{fig7.eps}}\bigg|_{k=0, \mu}\right)  .
\end{equation}
Now using Eq. (\ref{9}) in conjumination with Eq. (\ref{A43}), we find
\begin{eqnarray}\label{32}
B_{3} &=& \left(\frac{(N+2)(N+8)}{108}\right) \mu^{-3\epsilon}
\Bigl\{ - \frac{P^{2}}
{6\epsilon^{2}}\Bigr[1+\frac{1}{4}\epsilon-2\epsilon
\tilde{L}_{3}(P,\mu)\Bigr] \nonumber\\
&& - \frac{5\mu^{2}}{2\epsilon} \tilde{I}(P) \Bigr\}.
\end{eqnarray} 
\par Before proceeding, let us define the dimensionless bare couplings and 
the renormalization functions in minimal subtraction as powers series in the 
renormalized dimensionless coupling constant in the form
\begin{subequations}\label{33}
\begin{eqnarray}
&& u_{0} = u[1 + \sum_{i=1}^{\infty} a_{i}(\epsilon)
u^{i}], \label{33a}\\
&& Z_{\phi} = 1 + \sum_{i=1}^{\infty} b_{i}(\epsilon)
u^{i}, \label{33b}\\
&& \bar{Z}_{\phi^{2}} = 1 + \sum_{i=1}^{\infty} c_{i}(\epsilon)u^{i}.\label{33c}
\end{eqnarray}
\end{subequations}
By requiring minimal subtraction of dimensional poles, the renormalized 
primitively divergent vertex parts should be finite order by order in powers 
of $u$. This in turn determines $a_{i}(\epsilon), b_{i}(\epsilon)$ and 
$c_{i}(\epsilon)$. For the sake of simplicity, consider the renormalized vertices 
\begin{subequations}\label{34}
\begin{eqnarray}
&& \Gamma_{R}^{(2)}(k, u, m) = Z_{\phi}
\Gamma^{(2)}(k, u_{0},\mu), \label{34a}\\
&& \Gamma_{R}^{(4)}(k_{i}, u, m) = Z_{\phi}^{2} \Gamma^{(4)}(k_{i}, u_{0}, \mu), \label{34b}\\
&& \Gamma_{R}^{(2,1)}(k_{1}, k_{2}, p; u, m) = \bar{Z}_{\phi^{2}}
\Gamma^{(2,1)}(k_{1}, k_{2}, p, u_{0},\mu), \label{34c}
\end{eqnarray}
\end{subequations}
up to two-loop order (neglect the $B_{3}$ coefficient in the bare vertex 
$\Gamma^{(2)}(k, u,\mu)$). First, replace the expansion for $Z_{\phi}$ into 
$\Gamma_{R}^{(2)}(k, u, m)$ and define $m^{2} \equiv  Z_{\phi}\mu^{2}$ in the 
mass term which does not multiply coupling constant factors. Next, the value 
$u_{0}=u$ can surely be taken at this order. Recalling that regular terms are 
not taken into account in this set of steps, we find
\begin{eqnarray}\label{35}
&& \Gamma_{R}^{(2)}(k, u, m) = k^{2} + m^{2} + k^{2} b_{1}u 
+ k^{2}u^{2} \Bigr(b_{2}+ \frac{(N+2)}{144\epsilon}\Bigl).
\end{eqnarray}
The absence of poles in $\epsilon$ requires that $b_{1}=0$, which is 
consistent with the absence of the tadpole graph and 
$b_{2}=-\frac{(N+2)}{144\epsilon}$.  
\par Focusing our attention now in $\Gamma^{(4)}(k, u_{0},\mu)$, the coefficients appearing in 
its bare counterpart can be written in terms of the integrals computed in Appendix 
A, namely
\begin{subequations}\label{36}
\begin{eqnarray}
&& A_{1} = \frac{(N+8)}{18}[ I_{2}(k_{1} + k_{2}) 
+  I_{2}(k_{1} + k_{3}) + I_{2}(k_{2} + k_{3})] ,\label{36a}\\
&& A_{2}^{(1)} = \frac{(N^{2} + 6N + 20)}{108}
[I_{2}^{2}(k_{1} + k_{2})
+  I_{2}^{2}(k_{1} + k_{3}) + I_{2}^{2}(k_{2} + k_{3})] ,\label{36b}\\
&& A_{2}^{(2)} = \frac{(5N + 22)}{54}
[I_{4}(k_{i}) + 5perms.]. \label{36c}
\end{eqnarray}
\end{subequations}
Since $Z_{\phi}^{2}= 1 + 2 b_{2} u^{2} + O(u^{4})$, 
$\Gamma_{R}^{(4)}(k_{i}, u, m)$ can be expressed in the form
\begin{eqnarray}\label{37}
&& \Gamma_{R}^{(4)}(k_{i}, u, m)= \mu^{\epsilon}(1+ 2 b_{2} u^{2})(u + a_{1}u^{2} + a_{2}u^{3})[1 - A_{1} \mu^{\epsilon} (u + a_{1}u^{2}) \nonumber\\
&&\;\; + (A_{2}^{(1)} + A_{2}^{(2)})\mu^{2 \epsilon} u^{2}],
\end{eqnarray}
which can be simplified by grouping the powers of $u$ together. This leads to 
\begin{eqnarray}\label{38}
&& \Gamma_{R}^{(4)}(k_{i}, u, m)= \mu^{\epsilon}(u + 
(a_{1}- A_{1} \mu^{\epsilon})u^{2} + (a_{2} + 2 b_{2} - 2a_{1}A_{1} \mu^{\epsilon} 
\nonumber\\
&&\;\;\; + (A_{2}^{(1)} + A_{2}^{(2)})\mu^{2 \epsilon}) u^{3}).
\end{eqnarray}
By demanding that the poles be minimally cancelled at $O(u^{3})$, we employ 
Eqs. (\ref{A4}), (\ref{A5}), (\ref{A7}) and (\ref{A13}) combined with 
Eqs. (\ref{36}). In the resulting expression, all integrals 
$\tilde{L}(P)$ which appear in the several loop contributions of the 
four-point vertex function vanish and we obtain the 
following singular coefficients:
\begin{subequations}
\begin{eqnarray}\label{39}
&& a_{1} = \frac{(N + 8)}{6\epsilon}, \label{39a}\\
&& a_{2} = \frac{(N + 8)^{2}}{36\epsilon^{2}} - \frac{(3N + 14)}{24\epsilon}.\label{39b}
\end{eqnarray}
\end{subequations}
Let us complete our task at two-loop level by analyzing the vertex part 
$\Gamma^{(2,1)}$ in the computation of $\bar{Z}_{\phi^{2}}$. When we utilize Eqs. (\ref{21})-(\ref{24}) together with Eq. (\ref{26}), we can 
identify the coefficients present in Eq. (\ref{29c}) as 
\begin{subequations}
\begin{eqnarray}\label{40}
&& C_{1} = \frac{N+2}{18}[ I_{2}(k_{1} + k_{2}) +  I_{2}(k_{1} + k_{3}) 
+ I_{2}(k_{2} + k_{3})] ,\label{40a}\\
&& C_{2}^{(1)} = \frac{(N+2)^{2}}{108}
[I_{2}^{2}(k_{1} + k_{2})
+  I_{2}^{2}(k_{1} + k_{3}) + I_{2}^{2}(k_{2} + k_{3})] ,\label{40b}\\
&& C_{2}^{(2)} = \frac{N+2}{36}[I_{4}(k_{i}) + 5perms.].\label{40c}
\end{eqnarray}
\end{subequations}
Employing Eq. (\ref{34c}) in conjunction with the expansions of 
$\bar{Z}_{\phi^{2}}$ and $u_{0}$ in powers of $u$, we are led to 
\begin{eqnarray}\label{41}
&& \Gamma^{(2,1)}(k_{1}, k_{2}; p,u_{0}, \mu) = 1 + (c_{1}- 
C_{1} \mu^{\epsilon}) u + (c_{2}- (c_{1}+a_{1})\mu^{\epsilon}C_{1} \nonumber\\
&& \qquad   + \;\;(C_{2}^{(1)} + C_{2}^{(2)}) \mu^{2 \epsilon}) u^{2}.
\end{eqnarray}
Requirement of minimal cancellations of the poles allows to compute $c_{1}$ 
and $c_{2}$. Indeed, the integral $\tilde{L}(P)$ attached to 
$I_{2}$ and $I_{4}$ (see Appendix A) cancels out similarly to what took place 
in the renormalization of the $\Gamma^{(4)}$ vertex part, resulting in the 
coefficients:
\begin{subequations}
\begin{eqnarray}\label{42}
&& c_{1} = \frac{(N + 2)}{6\epsilon}, \label{42a}\\
&& c_{2} = \frac{(N + 2)(N + 5)}{36\epsilon^{2}} - \frac{(N + 2)}{24\epsilon}.\label{42b}
\end{eqnarray}
\end{subequations}  
\par So far everything is entirely similar to what happens in the minimal 
subtraction scheme for massless fields. 
\par Let us turn now our attention to the computation of $Z_{\phi}$ at 
three-loop level, i.e., we have to compute $b_{3}$ by using the diagrammatic 
expansion considering up to the $B_{3}$ contribution in $\Gamma^{(2)}$. 
Performing all the steps just like before, we have to be careful to use 
$u_{0}^{2}=u^{2} + \frac{(N+8)}{3\epsilon}u^{3}$. This produces a nontrivial mixing 
involving the $B_{2}$ and $B_{3}$ diagrams, which in the end of the day 
eliminates the $\tilde{L}_{3}(k)$ integrals appearing in both terms. Notice 
that $b_{3}$ is determined from the coefficient of the $k^{2}$ term. Working 
out the details, we find 
\begin{equation}\label{43}
b_{3} = - \frac{(N+2)(N+8)}{1296\epsilon^{2}} 
+ \frac{(N+2)(N+8)}{5184\epsilon}.
\end{equation}
However, this is not sufficient to subtract all the poles in $\epsilon$, since 
the resulting vertex part at this order is given by
\begin{eqnarray}\label{44}
&& \Gamma_{R}^{(2)}(k, u, m) = k^{2} + m^{2}\Bigl[1 + \frac{(N+2)}{24}
u^{2} \tilde{I}(k) \nonumber\\ 
&& - \frac{(N+2)(N+8)}{\epsilon}u^{3} \Bigl(\frac{\tilde{I}(k)}{108}\Bigr) \Bigr],
\end{eqnarray}
where from the Appendix A we know that
\begin{eqnarray}\label{45}
&& \tilde{I}(k) = \int_{0}^{1} dx \int_{0}^{1}dy lny 
\frac{d}{dy}\Bigl((1-y)ln\Bigl[\frac{y(1-y)\frac{k^{2}}{\mu^{2}} + 1-y 
+ \frac{y}{x(1-x)}}{1-y 
+ \frac{y}{x(1-x)}} \Bigr]\Bigr).
\end{eqnarray}
\par Although the integral is well behaved, we seem to be in trouble here, 
since the advertised minimal subtraction has produced a residual pole at 
three-loop level in contradiction with the minimal subtraction assertion in 
the first place! We already know that the theory is renormalizable by minimal 
subtraction had we used the 
tree-level bare mass $\mu_{0}$ and a larger number of diagrams. What is 
happening is obvious: the price to pay for a reparametrization 
$\mu_{0} \rightarrow \mu$ which eliminates many diagrams is the appearance of 
the above integral which seems to invalidate the minimal subtraction 
procedure only at three-loop order. 
\par The way out to this difficulty is the introduction of an extra subtraction in the two-point 
vertex function, in order to compensate for the bare mass reparametrization, 
which removes the remaining pole and we are done. Thus, by defining
\begin{eqnarray}\label{46}
&& \tilde{\Gamma}_{R}^{(2)}(k, u, m) = \Gamma_{R}^{(2)}(k, u, m) + m^{2}\Bigl[\frac{(N+2)(N+8)}{\epsilon} u^{3} \Bigl[\frac{\tilde{I}(k)}{108} \Bigr]  \Bigr],
\end{eqnarray}
we establish a direct connection with the minimal subtraction in the massless 
theory, since those terms proportional to $m^{2}$ are not there and this 
maneuver is unnecessary. In comparison with normalization conditions in the massive theory, at $k=0$ 
$\tilde{\Gamma}_{R}^{(2)}(0, u, m) = \Gamma_{R}^{(2)}(0, u, m)= m^{2}$ and we do not need to perform the extra subtraction.
\par Thence, our new proposal requires the vertex parts 
$\tilde{\Gamma}_{R}^{(2)}(k, u, m), \Gamma_{R}^{(4)}(k_{i}, u, m)$ and 
$\Gamma_{R}^{(2,1)}(k_{1}, k_{2}, p; u, m)$ to be renormalized by minimal 
subtraction with the same normalization functions  $Z_{\phi}$ and 
$\bar{Z}_{\phi^{2}}$ from the original renormalized field theory.
\par We are now in position to derive the exponents from our evaluation of 
the Wilson functions. In terms of the coefficients from  $Z_{\phi}, 
\bar{Z}_{\phi^{2}}$ and $u_{0}$, they are given by 
\begin{subequations}
\begin{eqnarray}\label{47}
&& \beta  =  -\epsilon u[1 - a_{1} u
+2(a_{1}^{2} -a_{2}) u^{2}], \label{47a}\\
&& \gamma_{\phi} = -\epsilon u [2b_{2} u
+ (3 b_{3}  - 2 b_{2} a_{1}) u^{2}], \label{47b}\\
&& \bar{\gamma}_{\phi^{2}} = \epsilon u [c_{1}
+ (2 c_{2}  - c_{1}^{2} - a_{1} c_{1})u    ]. \label{47c}
\end{eqnarray}
\end{subequations}
The eigenvalue condition $\beta(u_{\infty})= 0$ defines the repulsive fixed point 
$u_{\infty}$. It results in the following expression 
\begin{equation}\label{48}
u_{\infty}=\frac{6}{8 + N}\,\epsilon\Biggl\{1 + \epsilon
\,\frac{(9N + 42)}{(8 + N)^{2}}\Biggr\}\;\;.
\end{equation} 
Replacing this fixed point value we obtain the anomalous dimension of the 
field which is identical to the exponent $\eta\equiv \gamma_{\phi}(u_{\infty})$, namely
\begin{eqnarray}\label{49}
&&\eta= \frac{1}{2} \epsilon^{2}\,\frac{N + 2}{(N+8)^2}
\Bigl[1 + \epsilon(\frac{6(3N + 14)}{(N + 8)^{2}} - \frac{1}{4})\Bigr],
\end{eqnarray}
whereas the quantities $\gamma_{\phi^{2}}(u_{\infty})$ and 
$\bar{\gamma}_{\phi^{2}}(u_{\infty})$ are given by the 
expressions
\begin{subequations}
\begin{eqnarray}\label{50}
&& \gamma_{\phi^{2}}(u_{\infty})= \frac{N + 2}{(N+8)}\epsilon \Bigl[1 + 
\frac{(13N + 44)}{2(N+8)^{2}}\epsilon \Bigr], \label{50a}\\
&& \bar{\gamma}_{\phi^{2}}(u_{\infty}) = \frac{N + 2}{(N+8)}\epsilon \Bigl[1 + 
\frac{6(N + 3)}{(N+8)^{2}}\epsilon \Bigr]. \label{50b}
\end{eqnarray}
\end{subequations}
They are related to the anomalous dimension of the composite operator, also 
known as the correlation length exponent $\nu$, through the expression 
$\nu^{-1}= - d_{\phi^{2}}= 2 - \bar{\gamma}_{\phi^{2}}(u_{\infty})
- \gamma_{\phi}(u_{\infty})(= 2 -\gamma_{\phi^{2}}(u_{\infty}))$. 
Consequently, we can read off its value as  
\begin{eqnarray}\label{51}
&& \nu =\frac{1}{2} + \frac{(N + 2)}{4(N + 8)} \epsilon
+  \frac{1}{8}\frac{(N + 2)(N^{2} + 23N + 60)} {(N + 8)^3} \epsilon^{2}.
\end{eqnarray}
\par These critical exponents correspond to those previously found using: $i)$ massless 
fields within either the minimal subtraction or the normalization conditions formulation and $ii)$ massive 
theory using normalization conditions.  
\par The aforementioned results show that in spite of the extra subtraction 
that has to be carried over the vertex $\Gamma_{R}^{(2)}$ in our new 
procedure, the exponents are really the same when we use a minimal set of 
Feynman graphs in the determination of the renormalization functions. This minimal subtraction can be viewed as the counterpart 
of the normalization conditions framework $ii)$ from \cite{BLZ}. The relationship between the two massive formulations is entirely 
analogous to that shared by massless fields in $i)$. 
\par In order 
to compare with another standard method of minimal subtraction, we shall turn 
our attention to the BPHZ method in the next section which works with a much 
larger set of diagrams.
\section{BPHZ in minimal subtraction}
\par The BPH method involves the iterative introduction of counterterms in 
the original bare Lagrangian density in order to renormalize all vertex 
functions. These counterterms are required to obey the symmetries of the 
original bare Lagrangian. A very good account of this subject can be found in 
the book by Kleinert and Schulte-Frohlinde \cite{SFK}. 
\par We shall follow that material hereafter but point out some different conventions adopted in the present work. 
First, the symbol $\mathcal{K}()$ to be appplied in a diagram $()$ in order to pick out the singular part is going to be 
replaced by $()_{S}$. We only utilize this notation for one-loop diagrams and tadpoles insertions on them. In the two- and three-loop 
diagrams of the two-point function, we include some regular parts depending on logarithmic integrals in order to show how they cancel out in the renormalization algorithm. The other higher-loop diagrams 
will be displayed only with their singular parts. In this sense our operationalization of the method in the present section resembles 
the approach described in (chapter 9 of) that book within the spirit of Refs. \cite{BP,H}. Second, since we are going to restrict ourselves to a fixed loop order 
where we already know all the diagrams involved, we shall not make use of the formal $R$-operation which permits the generation of 
counterterms at arbitrary order in perturbation theory, albeit this procedure can also be applied and shown to be equivalent to 
the $BPH$ construction \cite{Z1}. Thus, we found appropriate to call this technique 
$BPHZ$ method. Third, the multiplicative factor appearing in each loop 
integral in the conventions of Kleinert and Schulte-Frohlinde's book is different from the conventions adopted 
here (see \cite{Amit} for a similar notation as ours). 
\par We shall maintain the 
notation presented so far, performing some adaptations from the method 
discussed in that book. In order to do so, we start with a quick description of 
the script necessary to carry out the computation of the critical exponents.
\par Using the bare Lagrangian density Eq. (\ref{1}) and 
performing the redefinitions $\phi_{0}=Z_{\phi}^{\frac{1}{2}}\phi$, 
$\mu_{0}^{2}=m^{2}\frac{Z_{m^{2}}}{Z_{\phi}}$ and 
$\lambda=\frac{Z_{u}}{Z_{\phi}^{2}} \mu^{\epsilon} u$, it turns out to be given by
\begin{equation}\label{52}
\mathcal{L} = \frac{1}{2}Z_{\phi}
|\bigtriangledown \phi|^{2} + \frac{1}{2} m^{2}Z_{m^{2}}\phi^{2} + \frac{1}{4!}\mu^{\epsilon}u Z_{u}(\phi^{2})^{2} ,
\end{equation}
whose coefficients are $\epsilon$-dependent. Note that the mass scale $\mu$ 
is arbitrary, $m$ is the renormalized mass, $u$ is the renormalized 
dimensionless coupling constant and $Z_{\phi}= 1 + \delta_{\phi}, 
Z_{m^{2}}= 1 + \delta_{m^{2}}$ and $Z_{u}= 1 + \delta_{u}$ are the 
renormalization functions. The amounts $\delta_{\phi}, \delta_{m^{2}}$ and 
$\delta_{u}$ are the counterterms which are added at each diagram in arbitrary 
loop order in order to cancel the singular contributions of the 
primitively divergent bare vertex parts. Denote the external momentum by $P$. 
The counterterms generate additional vertices and originate the following 
Feynman rules in momentum space 
\begin{subequations}
\begin{eqnarray}\label{53}
&& \parbox{12mm}{\includegraphics[scale=1.0]{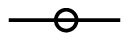}}  =  P^{2}\delta_{\phi}, \label{53a}\\
&& \parbox{15mm}{\includegraphics[scale=1.0]{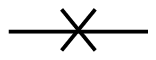}}   =  m^{2}\delta_{m^{2}}, \label{53b}\\
&& \parbox{8mm}{\includegraphics[scale=.1]{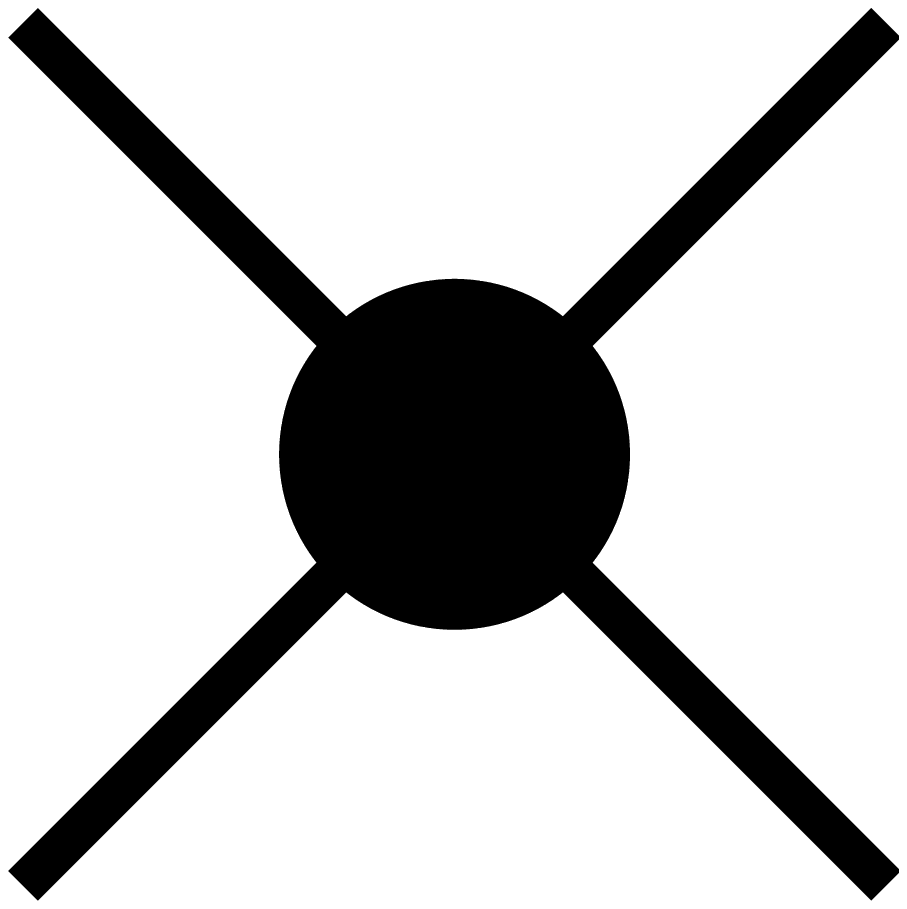}}  =  \mu^{\epsilon}u \delta_{u}. \label{53c}
\end{eqnarray}
\end{subequations}
In practice we shall need these counterterms expanded in powers of $u$ up to 
the desired order. In the present work, the counterterms will be expanded 
in the following form:
$\delta_{\phi}= \delta_{\phi}^{(1)}u + \delta_{\phi}^{(2)}u^{2} 
+ \delta_{\phi}^{(3)}u^{3}$, $\delta_{m^{2}}= \delta_{m^{2}}^{(1)}u 
+ \delta_{m^{2}}^{(2)}u^{2}$ and $\delta_{u}= \delta_{u}^{(1)}u 
+ \delta_{u}^{(2)}u^{2}$.   
\par In the present section 
we shall not need to determine the higher loop two-point function graphs in the degree of detail presented in Appendix A. As it 
is going to be shown in the remainder, only a simplified form of that vertex part suitable to our purposes will be worked out explicitly.
Every graph in the present
method is computed with the renormalized mass, but all integrals appearing 
depend on dimensionless ratios such as $\frac{k^{2}}{\mu^{2}}$ or 
$\frac{m^{2}}{\mu^{2}}$ which always show up in intermediate steps in the 
argument of logarithms. At least in a particular loop order, we shall verify that all of them cancel in the end 
of a sample calculation.
\par We start with the tadpole diagram. The corresponding integral is
\begin{eqnarray}\label{54}
I_{T} & = & \int d^{d}q \frac{1}{q^{2}+m^ {2}}.
\end{eqnarray}
Using Eq. (\ref{A2}), absorbing $S_{d}$ in the redefinition of the 
coupling constant, performing the continuation $d=4-\epsilon$ and the 
expansion up to $O(\epsilon^{0})$ we find 
\begin{eqnarray}\label{55}
I_{T} & = & -\frac{m^ {2}}{\epsilon}\left[1-\frac{\epsilon}{2}\ln(m^
{2})\right].
\end{eqnarray}
Therefore, when we include the $O(N)$ factor, the tadpole diagram is given 
by the following expression 
\begin{eqnarray}\label{56}
% tadpole
\parbox{10mm}{\includegraphics[scale=1.0]{fig1.eps}} & = & - \frac{(N+2)}{3} \frac{m^ {2}}{\epsilon}\left[1-\frac{\epsilon}{2}\ln(m^
{2})\right].
\end{eqnarray}
\par The double tadpole is characterized by the integral
\begin{eqnarray}\label{57}
I_{DT} & = & \int d^{d}q_{1}d^{d}q_{2} \frac{1}{(q_{1}^{2}+m^
{2})^{2}(q_{2}^{2}+m^ {2})}.
\end{eqnarray}
The integrals over $q_{1}$ and $q_{2}$ can be performed independently. The 
former is identical to a four-point one-loop diagram at zero external momenta, 
whereas the latter is given by Eq. (\ref{56}) discussed above. Using these 
facts, one can show that the double tadpole graph is represented by the expression
\begin{eqnarray}\label{58}
\parbox{6mm}{\includegraphics[scale=1.0]{fig2.eps}} \quad &=& - \left(\frac{N+2}{3} \right)^{2} 
\frac{m^
{2}}{\epsilon^{2}}\left[1-\frac{\epsilon}{2}-\epsilon\ln(m^ {2})\right].
\end{eqnarray}
\par The integral corresponding to the sunset diagram was already computed 
in Appendix A. In the notation of the present section it is given by 
\begin{eqnarray}
I_{3}(P,m)  & = & \int d^{d}q_{1}d^{d}q_{2} \frac{1}{(q_{1}^{2}+m^
{2})(q_{2}^{2}+m^ {2})[(q_{1}+q_{2}+P)^{2}+m^ {2}]}.\nonumber
\end{eqnarray}
\par When we employ a simpler form of its calculation sketched in Appendix A multiplied to the $O(N)$ factor, the sunset symbol 
is equivalent to the expression
\begin{eqnarray}\label{60}
% sunset
\parbox{8mm}{\includegraphics[scale=1.0]{fig6.eps}}  \quad &=& - \left(\frac{N+2}{3} \right)\left(\frac{3m^
{2}}{2\epsilon^{2}}\left[1+\frac{1}{2}\epsilon-\epsilon\ln(m^ {2})\right] +
\frac{P^{2}}{8\epsilon}\left[1+\frac{1}{4}\epsilon-2\epsilon L_{3}(P)\right]
\right), 
\end{eqnarray} 
where 
\begin{eqnarray}\label{59}
L_{3}(P) & = &
\int_{0}^{1}dxdy(1-y)\ln\left\{y(1-y)P^{2}+\left[1-y+\frac{y}{x(1-x)}\right]m^
{2}\right\}.
\end{eqnarray}
\par The three-loop contribution to the two-point function 
\begin{eqnarray}
I_{5}(P,m) & = & \int d^{d}q_{1}d^{d}q_{2}d^{d}q_{3}\frac{1}{(q_{1}^{2}+m^
{2})(q_{2}^{2}+m^ {2})(q_{3}^{2}+m^ {2})}\times\nonumber \\ & &
\frac{1}{[(q_{1}+q_{2}+P)^{2}+m^ {2}][(q_{1}+q_{3}+P)^{2}+m^ {2}]},\nonumber
\end{eqnarray}
was also computed in the Appendix A. Its solution in a form useful to our 
purposes in the present section conjugated to the symmetry factor associated to the $O(N)$ symmetry implies that 
the associated graph is given by 
\begin{eqnarray}\label{61}
\parbox{12mm}{\includegraphics[scale=1.0]{fig7.eps}} &=& - \frac{(N+2)(N+8)}{27}\Bigl(\frac{5m^
{2}}{3\epsilon^{3}}\Bigl[1+\epsilon \Bigl(1-\frac{3}{2}\ln(m^ {2})\Bigr) + \epsilon^{2} \Bigl(\frac{\pi^{2}}{24} + \frac{15}{4} 
+ \frac{9}{8}(ln(m)^{2})^{2} \nonumber\\
&& \; + \; \frac{3}{2} \tilde{i}(P)\Bigr)\Bigr] \; + \; \frac{P^{2}}{6\epsilon^{2}}\Bigl[1+\frac{1}{2}\epsilon-3\epsilon L_{3}(P)
\Bigr]\Bigr).
\end{eqnarray}
Since the last expression is $O(u^{3})$, the coefficient of the $\epsilon^{-3}$ part 
which is proportional to the mass only contribute to the mass countertem at 
$O(u^{3})$ which is of an order higher than we need in our present discussion. The combination of this term with those coming from 
three-loop tadpole diagrams is certainly important in the proof of mass renormalization at three-loops. We shall neglect it 
consistently with the arguments to be explained next.  
\par The remaining three-loop diagrams from the two-point vertex can be separated in two distinct sets. The first one corresponds 
to tadpole and four-point insertions into tadpole diagrams (Eqs. (\ref{5})-(\ref{7})) and do not 
depend upon the external momenta. Together with their 
counterterms, their singular part will contribute to the mass renormalization 
at three-loop level. Remember that we are interested only in the computation of $Z_{\phi}$ 
(proportional to $P^{2}$) up to three-loop order. We do not have to consider those diagrams for they will not 
contribute to the evaluation of $Z_{\phi}$. The second set corresponds solely 
to the ``sunset'' diagram with a tadpole insertion Eq. (\ref{10}) and its 
counterterm. In the Appendix B, we show explicitly that the singular 
parts (poles in $\epsilon$) coming from these integrals do not depend on the 
external momenta and also do not contribute to the computation of $Z_{\phi}$. We are going to consider them implicitly in 
the three-loop expansion of the two-point vertex part but shall not work them out from now on. They are going to be collectively referred 
to as ``tadpoles'' in the remainder of this section.
\par Next we shall analyze the graphs contributing to the four-point function. 
The one-loop integral 
\begin{eqnarray}
 I_{2}(P,m) & = & \int d^{d}q\frac{1}{(q^{2}+m^ {2})[(q+P)^{2}+m^ {2}]},
\nonumber
\end{eqnarray}
can be read off from Eq. (\ref{A4}) from Appendix A, which in conjuminance with Eq. (\ref{15}) produces the following result to its 
corresponding diagram: 
\begin{eqnarray}\label{63}
 \parbox{10mm}{\includegraphics[scale=1.0]{fig10.eps}} & = & \frac{(N+8)}{9} \frac{1}{\epsilon}\left[1-\frac{1}{2}\epsilon-\frac{1}{2}\epsilon L(P)\right],
\end{eqnarray}
where
\begin{eqnarray}\label{62}
L(P) & = & \int_{0}^{1}dx\ln[x(1-x)P^{2}+m^ {2}].
\end{eqnarray}
\par The integral associated to the diagram $\parbox{16mm}{\includegraphics[scale=1.0]{fig11.eps}}$ and  given by
%##########################################################################################################
\begin{eqnarray}
      \int d^{d}q_{1}d^{d}q_{2}\frac{1}{(q_{1}^{2}+m^
{2})[(q_{1}+P)^{2}+m^ {2}]} \frac{1}{(q_{2}^{2}+m^ {2})[(q_{2}+P)^{2}+m^ {2}]},
\nonumber
\end{eqnarray}
can be written diagrammatically as the square of the previous one-loop contribution. Therefore, we 
obtain the following expression
\begin{eqnarray}\label{64}
     \parbox{16mm}{\includegraphics[scale=1.0]{fig11.eps}} & = & \frac{(N^{2}+6N+20)}{27}\frac{1}{\epsilon^{2}}[1-\epsilon-\epsilon L(P)].
\end{eqnarray}
\par Consider the nontrivial two-loop contribution
\begin{eqnarray}
I_{4}(P,m) & = & \int d^{d}q_{1}d^{d}q_{2}\frac{1}{(q_{1}^{2}+m^
{2})[(P-q_{1})^{2}+m^ {2}](q_{2}^{2}+m^ {2})[(q_{1}-q_{2}+P_{3})^{2}+m^ {2}]}.
\nonumber
\end{eqnarray}
In order to compare with Eq. (\ref{A13}), we just have to replace the 
bare mass by the renormalized one $m$ without factoring it out from the 
integral. It then follows that the solution to its diagram can be written as
\begin{eqnarray}\label{65}
  \parbox{14mm}{\includegraphics[scale=1.0]{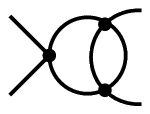}}   & = & \left(\frac{5N+22}{27}\right)\frac{1}{2\epsilon^{2}}\left[1-\frac{1}{2}\epsilon-\epsilon L(P)\right]. 
\end{eqnarray}
\par Another two-loop diagram contributing to the four-point vertex part is 
$\left(\parbox{10mm}{\includegraphics[scale=1.0]{fig13.eps}}\right)$, which is represented by the integral       
\begin{eqnarray}
I_{2T}(P,m) & = & \int d^{d}q_{1}d^{d}q_{2}\frac{1}{(q_{1}^{2}+m^
{2})^{2}[(q_{1}+P)^{2}+m^ {2}](q_{2}^{2}+m^ {2})}.\nonumber
\end{eqnarray}
The integral over $q_{2}$ is simply a tadpole, whereas the integral over $q_{1}$ can be evaluated using Feynman parameters. Only the 
singular part of its diagram will be interesting to our purposes and turns out to be  
\begin{eqnarray}\label{67}
\left[\parbox{10mm}{\includegraphics[scale=1.0]{fig13.eps}}\right]_{S} &=&- \frac{(N+2)(N+8)}{27}\left[\frac{m^{2}}{2\epsilon}\int_{0}^{1}dx\frac{(1-x)}{x(1-x)P^{2}+m^{2}}\right] 
. 
\end{eqnarray}
\par The method from last section considered the 
diagrammatic expansion without counterterms and the renormalization functions 
were obtained by demanding finite renormalized vertex functions. Here, the 
normalization functions are obtained directly from the counterterms generated 
order by order in perturbation theory. We start by using the 
diagrammatic expansion of the two-point vertex function up to two-loop order, 
which including counterms diagrams, reads
\begin{eqnarray}\label{68}
&& \Gamma^{(2)}(P, m,\mu^{\epsilon}u)= P^{2} + m^{2} + u 
\left(\frac{\mu^{\epsilon}}{2} 
\parbox{10mm}{\includegraphics[scale=1.0]{fig1.eps}} + m^{2}\delta_{m^{2}}^{(1)} + P^{2}\delta_{\phi}^{(1)}\right)
+ u^{2}\Bigl( - \quad \frac{\mu^{2\epsilon}}{4}
% tadpole duplo
\parbox{10mm}{\includegraphics[scale=1.0]{fig2.eps}} \nonumber \\
&& \quad - \quad \frac{\mu^{2\epsilon}}{6}
% sunset
\parbox{12mm}{\includegraphics[scale=1.0]{fig6.eps}}  - \frac{\mu^{\epsilon}m^ {2}\tilde{\lambda}_{m^ {2}}}{2u}
\parbox{6mm}{\includegraphics[scale=1.0]{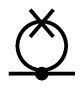}} \quad + \frac{\mu^{\epsilon}\tilde{\lambda}_{u}}{2u}
\parbox{10mm}{\includegraphics[scale=1.0]{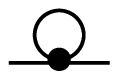}}  +  m^{2}\delta_{m^{2}}^{(2)} + P^{2}\delta_{\phi}^{(2)}\Bigr), 
\end{eqnarray}
where the last two diagrams are computed at zero external momentum since 
they are constructed out from tadpoles and shall be 
discussed in a moment. As the counterterms select just the singular part 
($\equiv ()_{S}$) of the diagrams, the conditions of finitenes of this vertex 
part at one-loop order ($O(u)$) are equivalent to the following identifications 
\begin{subequations}
\begin{eqnarray}\label{69}
&& m^{2}\delta_{m^{2}}^{(1)}= -\frac{\mu^{\epsilon}}{2}
\left( \parbox{10mm}{\includegraphics[scale=1.0]{fig1.eps}} \right)_{S},  \label{69a}\\
&& P^{2}\delta_{\phi}^{(1)}= -P^{2}\frac{\mu^{\epsilon}}{2}
\left( \parbox{10mm}{\includegraphics[scale=1.0]{fig1.eps}} \right)_{S},  \label{69b}
\end{eqnarray}
\end{subequations}
which in conjuminance with Eq. (\ref{56}), lead to the following coefficients 
\begin{subequations} \label{70}
\begin{eqnarray}
&& \delta_{m^{2}}^{(1)}=\frac{(N+2)}{6\epsilon},  \label{70a}\\
&& \delta_{\phi}^{(1)}=0. \label{70b}
\end{eqnarray}
\end{subequations}
Now consider the fourth graph in Eq. (\ref{68}). It is a double tadpole where 
the upper tadpole was replaced by its counterterm. In other 
words, the upper counterterm coupling constant can be identified through 
the relation $\tilde{\lambda}_{m^ {2}} \equiv 
u \delta_{m^{2}}^{(1)}$ and the diagram can be expressed in the form:
\begin{eqnarray}\label{71}
\frac{\mu^{\epsilon} m^{2} \tilde{\lambda}_{m^ {2}}}{2u}
\parbox{6mm}{\includegraphics[scale=1.0]{fig22.eps}}  &=& m^{2}\frac{(N+2)^{2}}{36\epsilon^{2}}\left[1-\frac{1}{2}\epsilon
-\frac{1}{2}\epsilon \ln\left(\frac{m^ {2}}{\mu^{2}}\right)\right].
\end{eqnarray}
\par Before going ahead, consider the one-loop 4-point vertex function. Its 
diagrammatic expansion including the counterterm is given by 
%Função de 4 Pontos%
\begin{eqnarray}\label{72}
&&\Gamma^{(4)}(k_{i},m,\mu^{\epsilon}u) =  u\mu^{\epsilon}\left(1 - u
\frac{\mu^{\epsilon}}{2}
 \Bigl(\Bigl[\parbox{10mm}{\includegraphics[scale=1.0]{fig10.eps}}\Bigr](k_{1}+k_{2}) + 2 \;permutations\Bigr) 
+ u\delta_{u}^{(1)}\right).
\end{eqnarray}
By using Eq. (\ref{63}), this vertex part is finite if the following relation 
holds:
\begin{eqnarray}\label{73}
 \delta_{u}^{(1)}= \frac{(N+8)}{6\epsilon}.
\end{eqnarray}
\par The fifth (last counterterm) diagram of the two-point 
function in Eq. (\ref{68}) is a product of a tadpole with a four-point 
insertion, where the latter loop has shrunken to zero but picking out the 
coupling constant from its singular counterterm. This is equivalent to take 
$\tilde{\lambda}_{u} \equiv u\delta_{u}^{(1)}$ and its corresponding 
expression reads:
\begin{eqnarray}\label{74}  
\frac{\mu^{\epsilon}\tilde{\lambda}_{u}}{2u}
\parbox{10mm}{\includegraphics[scale=1.0]{fig23.eps}}  &=& - m^{2}\frac{(N+2)(N+8)}{36\epsilon^{2}}\left[1-\frac{\epsilon}{2}\ln\left(\frac{m^{2}}{\mu^{2}}\right)\right].
\end{eqnarray}
Now we replace the results of this discussion in Eq. (\ref{68}) in order to 
determine $\delta_{m^{2}}^{(2)}$ and $\delta_{\phi}^{(2)}$ at two-loop 
level. Indeed, substitution of the Eqs. (\ref{58}), (\ref{60}), (\ref{71}) 
and (\ref{74}) into Eq. (\ref{68}) followed by the expansion in powers 
of $\mu^{n\epsilon}$ up to first order in $\epsilon$, we find that all the 
terms $ln\left(\frac{m^{2}}{\mu^{2}}\right)$ cancel out at $O(u^{2})$. We then 
obtain:
\begin{subequations}
\begin{eqnarray}\label{75}  
&& \delta_{m^{2}}^{(2)}=\frac{(N+2)(N+5)}{36\epsilon^{2}} - \frac{(N+2)}{24\epsilon}, \label{75a}  \\
&& \delta_{\phi}^{(2)}= - \frac{(N+2)}{144\epsilon}.\label{75b}  
\end{eqnarray}
\end{subequations}
\par Let us examine the two-loop contribution from $\Gamma^{(4)}$. We just have to focus on the two-loop diagrams in 
order to determine the counterterm $\delta_{u}^{(2)}$, namely 
\begin{eqnarray}\label{76}
&&\Gamma_{2-loop}^{(4)}(k_{i},m, \mu^{\epsilon}u) = \mu^{\epsilon}u^{3} 
\Bigl[ \frac{\mu^{2\epsilon}}{4}
         \Bigl(\Bigl[\parbox{16mm}{\includegraphics[scale=1.0]{fig11.eps}}\Bigr](k_{1}+k_{2}) + 2perms.\Bigr) \nonumber\\
&& + \quad \frac{\mu^{2\epsilon}}{2}
   \Bigl(\Bigl[\parbox{10mm}{\includegraphics[scale=1.0]{fig12.eps}}\Bigr](k_{i}) + 5perms. \Bigr)
+ \quad\frac{\mu^{2\epsilon}}{2}
  \Bigl(\Bigl[\parbox{11mm}{\includegraphics[scale=1.0]{fig13.eps}}\Bigr](k_{1}+k_{2}) + 2perms. \Bigr) \nonumber\\
&& + \;\frac{\mu^{\epsilon} m^{2} \tilde{\lambda}_{m^ {2}}}{2u}
\Bigl(\Bigl[\parbox{11mm}{\includegraphics[scale=1.0]{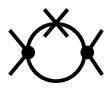}}\Bigr] (k_{1}+k_{2}) + 2perms. \Bigr)\Bigr)\;
- \frac{\mu^{\epsilon}\tilde{\lambda}_{u}}{u}\Bigl(\Bigl[\parbox{11mm}{\includegraphics[scale=1.0]{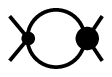}}\Bigr] (k_{1}+k_{2}) + 2perms. \Bigr) \nonumber\\ 
&& \;+ \;\delta_{u}^{(2)}\Bigr].
\end{eqnarray} 
The counterterm diagram corresponding to the fourth type of graphs appearing 
in  the last expansion is given by 
\begin{eqnarray}\label{77}
&& \Bigl[\parbox{11mm}{\includegraphics[scale=1.0]{fig24.eps}}\Bigr](k)  =  \frac{(N+8)}{9} \int d^{d}q_{1}\frac{1}{(q^{2}+m^
{2})^{2}[(q+k)^{2}+m^ {2}]}\nonumber\\
&& = \left[\frac{(N+8)}{18}\int_{0}^{1}dx\frac{(1-x)}{x(1-x)k^{2}+m^{2}}\right] + O(\epsilon). 
\end{eqnarray}
The multiplication of these diagrams by the factor $\frac{\mu^{\epsilon} m^{2} \tilde{\lambda}_{m^ {2}}}{2u}$ have a singular part which cancels exactly the 
contribution of the third kind of diagrams. What is left is the requirement 
that $\delta_{u}^{(2)}$ should subtract minimally the poles of the two-loop 
diagrams from $\Gamma_{2-loop}^{(4)}(k_{i},m, \mu^{\epsilon}u)$, i.e.,
\begin{eqnarray}\label{78}
&& \delta_{u}^{(2)} = \frac{\mu^{2\epsilon}}{4}
         \Bigl(\Bigl[\parbox{16mm}{\includegraphics[scale=1.0]{fig11.eps}}\Bigr](k_{1}+k_{2}) + 2 \;\; perms.\Bigr)_{S}  \;\; + \;\; \frac{\mu^{2\epsilon}}{2}
   \Bigl(\Bigl[\parbox{10mm}{\includegraphics[scale=1.0]{fig12.eps}}\Bigr](k_{i}) + 5 \;\;perms. \Bigr)_{S}\nonumber\\
&&\quad - \quad\frac{\mu^{\epsilon}\tilde{\lambda}_{u}}{u}\Bigl(\Bigl[\parbox{11mm}{\includegraphics[scale=1.0]{fig24.eps}}\Bigr] (k_{1}+k_{2}) + 2 \;\;perms. \Bigr)_{S}.
\end{eqnarray}  
The last diagram is just a one-loop diagram of the four-point coupling 
constant with its associated symmetry factor attached. When it is 
multiplied by $\frac{\mu^{\epsilon}\tilde{\lambda}_{u}}{u}$, with 
$\tilde{\lambda}_{u}=u\delta_{u}^{(1)}$, we find 
\begin{eqnarray}\label{79}
&& \frac{\mu^{\epsilon}\tilde{\lambda}_{u}}{u}\Bigl(\Bigl[\parbox{11mm}{\includegraphics[scale=1.0]{fig25.eps}}\Bigr] (k_{1}+k_{2})\Bigr)_{S} = \frac{(N+8)^{2}}{54\epsilon^{2}}
\left(1- \frac{\epsilon}{2} - \frac{\epsilon}{2}\hat{L}(k_{1}+k_{2})\right),
\end{eqnarray}
where 
\begin{eqnarray}\label{80}
\hat{L}(P) & = & \int_{0}^{1}dx\ln\left[\frac{x(1-x)P^{2}+m^ {2}}{\mu^{2}}\right].
\end{eqnarray}
 When we expand the factors $\mu^{n \epsilon}$ in powers of logarithms, the 
integral $L(P)$ in Eq. (\ref{62}) gets transformed to $\hat{L}(P)$ in all diagrams appearing in Eq. (\ref{78}). Summing up 
everything utilizing Eqs. (\ref{64}) and (\ref{65}) in conjunction with the last 
expressions, we verify that all terms proportional to $\hat{L}(P)$ with 
($P=k_{1}+k_{2},k_{1}+k_{3}$ and $k_{2}+k_{3}$) vanish. Therefore, this 
manipulation produces the following result
\begin{eqnarray}\label{81}
&& \delta_{u}^{(2)} =\frac{(N+8)^{2}}{36\epsilon^{2}} - 
\frac{(5N+22)}{36\epsilon}. 
\end{eqnarray}  
\par The three-loop diagrams of the two-point vertex part include ``tadpoles'' as well as relevant graphs to our computation of 
$\delta_{\phi}^{(3)}$. We can get rid of the former setting $m=0$ into their solution, which at the same time eliminates contributions 
to the mass renormalization at three-loops (proportional to $m^{2}$) and we do not have to worry about those terms at this perturbative 
order. For instance, the symbol 
$(diagram)_{m^{2}=0}$ implement the last condition. 
\par The three-loop diagrams of $\Gamma^{(2)}$, for the purpose of 
computing $\delta_{\phi}^{(3)}$, can be written in a simplified form as
\begin{eqnarray}\label{82}
\Gamma^{(2)}_{3-loop}(P) &=& u^{3}\Bigl(\frac{\mu^{3\epsilon}}{4}
\Bigl[\parbox{12mm}{\includegraphics[scale=1.0]{fig7.eps}}\Bigr]_{m^{2}=0} \; -  \;\frac{\mu^{2\epsilon}\tilde{\lambda}_{u}}{3u}
  \Bigl[\parbox{11mm}{\includegraphics[scale=1.0]{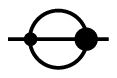}}\Bigr]_{m^{2}=0} \;+ P^{2}\delta_{\phi}^{(3)}\; +\; tadpoles \Bigr),
\end{eqnarray}
and from now on we are going to neglect the contributions coming from the 
tadpoles. Notice that even if in the remaining two diagrams the terms proportional to 
$m^{2}$ are going to be set to zero, we are not going to employ this simplification in the terms proportional to $P^{2}$, 
since we still have to demonstrate the elimination of $L_{3}(P)$-type 
contributions. Indeed, the object which appears when the $\mu^{n\epsilon}$ 
coefficient is expanded is given by:
\begin{eqnarray}\label{83}
\hat{L}_{3}(P) &=& \int_{0}^{1}dxdy(1-y)\ln\left\{\frac{y(1-y)P^{2}+\Bigl[1-y+\frac{y}{x(1-x)}\Bigr]m^
{2}}{\mu^{2}}\right\}.
\end{eqnarray}
Combining the last equation with the definition of $\tilde{\lambda}_{u}$, 
the solution of the diagrams represented by Eqs. (\ref{60}), (\ref{61}) and 
recalling the above remarks one can show that the $\hat{L}_{3}(P)$ 
contributions vanish. What remains after the cancellation of the dimensional poles is the identification of the normalization 
coefficient:
\begin{eqnarray}\label{84}
\delta_{\phi}^{(3)}= -\frac{(N+2)(N+8)}{1296\epsilon^{2}} \left(1 - 
\frac{\epsilon}{4}\right). 
\end{eqnarray}  
Therefore, the complete solution at the loop order required for the 
three normalization functions is represented by
\begin{subequations}\label{85}
\begin{eqnarray}
Z_{\phi}  &=& 1 - \frac{(N+2)}{144\epsilon}u^{2} -\frac{(N+2)(N+8)}{1296\epsilon^{2}} \left(1 - 
\frac{\epsilon}{4}\right)u^{3}, \label{85a}\\
Z_{m^{2}} &=& 1 + \frac{(N+2)}{6\epsilon}u + \left[\frac{(N+2)(N+5)}{36\epsilon^{2}} - \frac{(N+2)}{24\epsilon}\right]u^{2}, \label{85b}\\
Z_{u}     &=& 1 + \frac{(N+8)}{6\epsilon}u + \left[\frac{(N+8)^{2}}{36\epsilon^{2}} - 
\frac{(5N+22)}{36\epsilon}\right]u^{2}.\label{85c}
\end{eqnarray}
\end{subequations}
\par We have at hand the tools required to calculate the critical exponents. We start by the definition 
\begin{eqnarray}\label{86}
\beta(u)= \mu \left(\frac{\partial u}{\partial \mu}\right)_{[\mu_{0},\lambda]}= -\mu \left[\frac{(\frac{\partial \lambda}{\partial \mu})_{(\mu_{0},u)}}
{(\frac{\partial \lambda}{\partial u})_{(\mu_{0},\mu)}} \right].
\end{eqnarray}
Remember that $\lambda= Z_{u} Z_{\phi}^{-2} \mu^{\epsilon} u$. Then, it 
follows directly that
\begin{eqnarray}\label{87}
\beta(u)= -\epsilon \frac{\partial ln[Z_{u} Z_{\phi}^{-2}u]^{-1}}{\partial u} . \end{eqnarray}
Utilizing Eqs. (\ref{85a})-(\ref{85b}), we can rewrite last expression in terms of the 
coefficients just obtained after some algebra as 
\begin{eqnarray}\label{88}
&& \beta  =  -\epsilon u[1 - \delta _{u}^{(1)} u
+2((\delta _{u}^{1})^{2} - \delta _{u}^{(2)} + 2\delta _{\phi}^{(2)}) u^{2}].
\end{eqnarray}
It is important to mention that this expression is formally different from 
Eq. (\ref{47a}), since the individual terms in the latter after convenient 
identifications are not identical. Nevertheless, in terms of the explicit 
computations already performed last equation can be written as 
\begin{eqnarray}\label{89}
&& \beta (u) = -\epsilon u + \frac{(N+8)}{6}u^{2} - \frac{(3N+14)}{12}u^{3},
\end{eqnarray}
which is exactly the same result that would have been obtained from the 
explicit substitution of the coefficients into Eq. (\ref{47a}).
\par Next, define the quantity $\gamma_{\phi}(u) = \mu \left(\frac{\partial  ln Z_{\phi}}{\partial \mu}\right)_{[\mu_{0},\lambda]}= \beta(u)
\left(\frac{\partial ln Z_{\phi}}{\partial u}\right)$. In terms of the 
various coefficients, it is given by
\begin{eqnarray}\label{90}
\gamma_{\phi}(u)= -\epsilon u[2\delta _{\phi}^{(2)}u + (3\delta _{\phi}^{(3)}
-2\delta _{u}^{(1)}\delta _{\phi}^{(2)})u^{2}], 
\end{eqnarray}
which is equivalent to 
\begin{eqnarray}\label{91}
\gamma_{\phi}(u)= \frac{(N+2)}{72}u^{2} - \frac{(N+2)(N+8)}{1728}u^{3}.
\end{eqnarray} 
The last two equations are identical to those from our previous unconventional 
description.
Now we introduce the amount $\gamma_{m}(u)= \frac{\mu}{m^{2}}\left(\frac{\partial  m^{2}}{\partial \mu}\right)_{[\mu_{0},\lambda]}= \gamma_{\phi}(u) - 
\beta(u)\left(\frac{\partial  ln Z_{m^{2}}}{\partial u}\right)$. It is 
convenient also to employ the notation $\bar{\gamma}_{m}(u) = - 
\beta(u)\left(\frac{\partial  ln Z_{m^{2}}}{\partial u}\right)$. It is easy 
to show that 
\begin{eqnarray}\label{92}
\gamma_{m}(u)= \gamma_{\phi}(u) + \epsilon u[\delta_{m^{2}}^{(1)} + (2\delta_{m^{2}}^{(2)} - (\delta_{m^{2}}^{(1)})^{2} 
- \delta _{u}^{(1)}\delta_{m^{2}}^{(1)})u].
\end{eqnarray}
We can work this out further in order to obtain the simpler expression
\begin{eqnarray}\label{93}
\gamma_{m}(u)= \frac{(N+2)}{6}u\left[1 - \frac{5}{12}u\right].
\end{eqnarray}
The nontrivial fixed point is given by the eigenvalue condition 
$\beta(\tilde{u}_{\infty})=0$, namely
\begin{eqnarray}\label{94} 
\tilde{u}_{\infty}&=&\frac{6\epsilon}{(N+8)}\left[1 
+ \frac{3(3N+14)\epsilon}{(N+8)^{2}}\right].
\end{eqnarray}
It turns out that $\gamma_{\phi}(\tilde{u}_{\infty})$ is simply the anomalous 
dimension of the field, exponent $\eta$ from Eq. (\ref{49}), namely
\begin{eqnarray}\label{95}
&& \eta=\gamma_{\phi}(\tilde{u}_{\infty}).
\end{eqnarray} 
Moreover, at the fixed point we find out that
\begin{eqnarray}\label{96}
&& \gamma_{m}(\tilde{u}_{\infty})=  \frac{N + 2}{(N+8)}\epsilon\Bigl[1 + 
\frac{(13N + 44)}{2(N + 8)^{2}}\epsilon \Bigr].
\end{eqnarray}
Note that this expression is identical to Eq. (\ref{50a}) for the Wilson 
function of the composite field. The above definitions get transliterated in 
the previous unconventional minimal subtraction through the identifications 
$\gamma_{m}(u)=\gamma_{\phi^{2}}(u)$ and 
$\bar{\gamma}_{m}(u) = \bar{\gamma}_{\phi^{2}}(u)$. The exponent $\nu$ is 
related to $\gamma_{m}(\tilde{u}_{\infty})$ through the relation $\nu=(2-\gamma_{m}(\tilde{u}_{\infty}))^{-1}$. Consequently, it is simple to demonstrate 
that $\nu$ obtained from this expression is identical to the result from Eq. (\ref{51}).

\section{Discussion of the results and Conclusion}
\par It is worthy to mention that the universal results coming from the 
new unconventional subtraction method introduced in Secs. II and III and the 
BPHZ method discussed  in the Section IV are identical as expected, 
even though the intermediate steps are quite different. 
\par Rigorously speaking, the BPHZ does not require any regularization method, 
since it is designed to yield finite expressions from any diagram by 
subtracting the divergent part without specification to the regulator employed. We used dimensional regularization in order to 
compare the same technique with previous results obtained using different 
conventions \cite{SFK}. We restricted ourselves only to the (large number of) 
diagrams strictly necessary to perform the explicit calculations of all 
quantities required in the computation of the critical exponents at the 
desired order in perturbation theory.
\par The unconventional method, on the other hand, requires an extra 
subtraction for the two-point function beyond three-loop order due to our 
choice of using the three-loop bare mass instead of employing the tree-level 
bare mass. The advantage is that all the tadpoles diagrams do not need to be 
considered in this framework, since they drop out trivially after expanding 
the bare propagator in terms of the new bare mass inside all diagrams. 
\par It is interesting to notice that despite the extra subtraction above 
mentioned, we can relate the method rather simply to minimal subtraction in 
the massless theory and to the massive theory using normalization conditions 
at the same order in perturbation theory. In fact, since the extra subtraction 
in the new method is proportional to the renormalized mass, at zero mass this 
subtraction is identically zero. Of course, in that case the mass scale $\mu$ 
is replaced by a external nonvanishing momentum scale $\kappa$ and the 
cancellations involving integrals of logarithms of the external momentum in 
the massless case carry out in the same way as in the unconventional method. 
Second, we recover the normalization conditions for the massive theory at zero 
external momentum with a minimal number of diagrams as proposed in 
\cite{BLZ1}, which is also rather similar to the massless case.      
\par The choice of the three-loop bare mass to compute Feynman diagrams 
implies that only beyond three-loops we need the extra subtraction at the 
two-point vertex function, since everything works in exactly the same way at 
two-loop order whether we manipulate the minimal set of diagrams or if we 
utilize the full set of graphs. In particular, the extra subtraction involves 
an integral which is essentially different from the logarithimic integrals 
which also can multiply poles in $\epsilon$. The last integrals do not vanish 
at zero external momenta, while the new integral presented is identically 
zero at vanishing external momenta. Although it is certainly not polynomial 
in the external momenta, it does behave as a polynomial in the external momentum 
and can be tacitly identified with a ``harmless pole'' \cite{tHV} (which, 
rigorously speaking, is defined only when the residue of the pole is a 
polynomial of finite order in the external momenta) and the 
extra subtraction becomes natural within this context.
\par The consistency of the unconventional method is warranted when we 
confront it with its BPHZ standard counterpart. Along with the match 
of the field normalization constant in both formalisms, the identification of 
the composed field with the mass renormalization constant is exact. The beta 
functions are different in both frameworks but the combinations of 
the several diagrams produce the same answer: they yield the same fixed 
point. The other Wilson functions are proved to be the same in both schemes 
which lead to the same critical exponents. 
\par For higher loops we expect that 
more differences between these two methods can appear. A generic feature 
should be the appearance of more extra subtractions at the two-point function 
vertex part consisting of integrals that behave themselves as harmless poles. 
As discussed, this complication is directly connected with the definition of 
the bare mass at that loop order. On the other hand, the simplification 
achieved in the elimination of all tadpole insertions in the vertex parts 
required in the perturbative calculation of the critical exponents at that 
order compensates this extra subtraction. 
\par The present unconventional minimal subtraction procedure is completely different from traditional resummation 
methods designed to extract numerical estimations from the perturbative computation of physical quantities \cite{SFK}. In our approach, the normalization constants are obtained in the weak-coupling limit as explained above. The analysis in the remainder amounts to take the physical critical system at the (repulsive) fixed point keeping, however, a nonvanishing 
renormalized mass which prevents the system of going to the strong-coupling regime. It is also distinct in comparison with variational perturbation theory \cite{SFK,Klei}, for in that method the weak-coupling renormalization constants are 
obtained in the conventional way, the bare coupling constant is taken to infinity (renormalized mass tends to zero at the attractive massless fixed point) characterizing the strong-coupling regime where the resummation is then defined in $d= 4-\epsilon$. The comparison with three-dimensional systems could then be done by choosing either $\epsilon=1$ or by starting from scratch with fixed dimension $d=3$ and performing numerically the Feynman integrals and computing the critical exponents in the strong-coupling regime of small masses but without explicit scale invariance \cite{Com}. We emphasize that our unconventional resummation here just obtains the $\epsilon$-expansion results for the exponents. The aforementioned conventional resummations involving the $\epsilon$-expansion could be applied to our approach (involving 
the renormalized vertex parts $\tilde{\Gamma}_{R}^{(2)}$, $\Gamma_{R}^{(4)}$, $\Gamma_{R}^{(2,1)}$, and $u_{0}(u))$ since they represent an extra ingredient in ameliorating the convergence properties of the $\epsilon$-expansion as far as numerical estimations (e.g., of critical exponets) are concerned.
\par Applications of the unconventional as well as the standard BPHZ minimal 
subtraction methods could be employed in some problems involving the formulation of critical 
phenomena using massive scalar field theories. For instance, in calculating critical exponents or other universal 
quantities of ordinary finite size systems in a parallel plate layered geometry \cite{NF,SL}. In addition, the investigation 
of those techniques could shed new light in massive $\phi^{4}$ scalar theories when two mass 
scales are present as is the case in anisotropic $m$-axial Lifshitz 
criticalities \cite{CL1}. This is a natural extension of the formalism 
discussed in the present paper. Since the massive theory is more appealing in 
its connection with quantum field theory, it would be interesting to 
investigate the perturbative analysis concerning field theories in Lifshitz 
spacetimes (see, for example, Refs. \cite{Horava} and \cite{Anselmi}). This constitutes a nice 
prelude to the treatment of a similar problem with several mass scales 
appearing naturally in the context of anisotropic generic competing systems 
of the Lifshitz type \cite{CL2} and its future potential applications in 
quantum field theory.

\appendix
\section{Minimal set of massive integrals in dimensional regularization}
\par Here we compute the minimal set of integrals to be used in Sections III and IV. In the $BPHZ$ method, the results 
which are going to be demonstrated should be complemented with additional information which is the 
content of Appendix B. 
\par Since this computation is well known from textbooks, we 
shall try to reduce the number of steps in getting the solution of the 
integrals in an attempt to fix our conventions in a more or less 
self-contained form. In the new method, we just 
need to replace the tree-level bare mass $\mu_{0}$ by the three-loop bare 
mass $\mu$ in all integrals as explained in the main text. We shall switch 
to $m$ in the $BPHZ$ technique discussed in Appendix B and in 
the appropriate places in the body of the paper.
\par The integrals connected to the four-point function vertex part diagrams are the 
one-loop integral $I_{2}(k_{i})$ from Eq. (\ref{15}), the trivial 
two-loop contribution is denoted by $I_{2}^{2}(k_{i})$ (Eq. (\ref{16})) whereas 
the nontrivial two-loop correction is named $I_{4}(k_{i})$ (Eq. (\ref{17})), i.e., the first, second and third 
graphs of $\Gamma^{(4)}(k_{i})$ from Eq. (\ref{20}).    
\par The one-loop integral is given by
\begin{eqnarray}\label{A1}
I_{2}(P) & = & \int d^{d}q\frac{1}{(q^{2}+\mu^ {2})[(q+P)^{2}+\mu^
{2}]}.
\end{eqnarray}
We introduce a Feynman parameter $x$ and use the following useful identity 
in order to set the notation from Ref. \cite{Amit} in the computation of all 
diagrams  
\begin{equation}\label{A2}
\int \frac {d^{d}q}{(q^{2} + 2 k.q + m^{2})^{\alpha}} =
\frac{1}{2} \frac{\Gamma(\frac{d}{2}) \Gamma(\alpha - \frac{d}{2}) (m^{2} - k^{2})^{\frac{d}{2} - \alpha}}{\Gamma(\alpha)} S_{d},
\end{equation} 
where $S_{d}$ is the area of the $d$-dimensional unit sphere. After 
expanding $d=4-\epsilon$ in the argument of the $\Gamma$ function and using 
the property $\Gamma(1+z)=z \Gamma(z)$, this integral can be rewritten as 
\begin{eqnarray}\label{A3}
I_{2}(P) & = &
\frac{S_{d}}{\epsilon}\left(1-\frac{1}{2}\epsilon\right)\int_{0}^{1}dx[x(1-x)P^{2}+\mu^
{2}]^{-\epsilon/2}.
\end{eqnarray}
\par Note that everytime we perform a loop integral, the angular factor 
$S_{d}$ is included in the final answer of the integral. Our primitive vertex 
parts have a rather interesting property: the expansion in the number of loops 
actually coincides with an expansion in powers of the coupling constant, 
provided that we factor out the tree-level coupling constant of the four-point 
vertex part. Thus we can absorb the angular factor in the redefinition of the 
coupling constant. If we proceed in this way, this is equivalent to divide 
each loop integral performed by $S_{d}$ and this overall factor disappears in 
the final answer. We shall take this step into account hereafter in all loop 
integrals. Consequently, last integral becomes
\begin{eqnarray}\label{A4}
I_{2}(P) & = &
\frac{\mu^{-\epsilon}}{\epsilon}\left[1-\frac{1}{2}\epsilon-\frac{1}{2}\epsilon
\tilde{L}(P)\right],
\end{eqnarray}
where
\begin{eqnarray}\label{A5}
\tilde{L}(P) & = & \int_{0}^{1}dx\ln\Bigl[x(1-x)\frac{P^{2}}{\mu^ {2}} + 1\Bigr].
\end{eqnarray}
When integrals like this (see also $\tilde{L_{3}}$ below) are mutiplied by inverse powers of $\epsilon$ they must cancel in the 
renormalization algorithm.  
\par Since the diagrammatic identity is valid 
\begin{eqnarray}\label{A6}
     \parbox{15mm}{\includegraphics[scale=1.0]{fig11.eps}} & = & \left(
 \parbox{10mm}{\includegraphics[scale=1.0]{fig10.eps}}\right)^{2},
\end{eqnarray}
the integral corresponding to this diagram yields $I_{2}^{2}(P)$ and to the 
order required can be written in the form 
\begin{eqnarray}\label{A7}
I_{2}^{2}(P) & = & \frac{\mu^{-2\epsilon}}{\epsilon^{2}}[1-\epsilon-\epsilon 
\tilde{L}(P)].
\end{eqnarray}
In these integrals we recall that $P$ corresponds to the three possible 
combinations $k_{1}+k_{2}$, $k_{1}+k_{3}$ and $k_{2}+k_{3}$. 
\par Finally, one of the graphs pertaining to the nontrivial 2-loop 
contribution of the four-point function is given by 
\begin{eqnarray}\label{A8}
I_{4}(k_{i}) & = & \int d^{d}q_{1}d^{d}q_{2}\frac{1}{(q_{1}^{2}+\mu^
{2})[(P-q_{1})^{2}+\mu^ {2}](q_{2}^{2}+\mu^
{2})[(q_{1}-q_{2}+k_{3})^{2}+\mu^ {2}]},
\end{eqnarray}
where $P=k_{1}+k_{2}$. Although there are five more diagrams of this type 
contributing, we stick to this particular distribution of external momenta. 
After introducing a Feynman parameter and integrating over $q_{2}$ we find 
\begin{eqnarray}\label{A9}
I_{4}(k_{i})& = &
\frac{1}{\epsilon}\left(1-\frac{1}{2}\epsilon\right)\int_{0}^{1}dx[x(1-x)]^{-\epsilon/2}
\times\nonumber \\ & & \int d^{d}q_{1}\frac{1}{(q_{1}^{2}+\mu^
{2})[(P-q_{1})^{2}+\mu^ {2}][(q_{1}+P_{3})^{2}+m_{x}^{2}]^{\epsilon/2}},
\end{eqnarray}
where 
\begin{eqnarray}\label{A10}
m_{x}^{2}=\frac{\mu^ {2}}{x(1-x)}.
\end{eqnarray}
Using in sequence two Feynman parameters $z$ and $y$ and integrating over 
$q_{1}$, the integral takes the purely parametric form
\begin{eqnarray}\label{A11}
&&I_{4}(k_{i}) = 
\frac{1}{4\epsilon}(1-\epsilon)\int_{0}^{1}dx[x(1-x)]^{-\epsilon/2}\int_{0}^{1}dy(1-y)^{\epsilon/2-1}y
\times\nonumber \\ & &
 \int_{0}^{1}dz[yz(1-yz)P^{2}+y(1-y)P_{3}^{2}+2yz(1-y)P_{3}P+\mu^{2}y+m_{x}^{2}(1-y)]^{-\epsilon}.
\end{eqnarray}
Notice that the parametric integral is divergent at $y=1$ when 
$\epsilon=0$. In this diagram the leading divergences translate themselves 
as poles of the form $\frac{1}{\epsilon^{2}}$ and $\frac{1}{\epsilon}$. The 
term between brackets which multiply the $y$ integral can be written as  
$\{\}^{-\epsilon}= \{\}^{-\epsilon}|_{y=1} + [\{\}^{-\epsilon} 
- \{\}^{-\epsilon}|_{y=1}]$. Next, since $a^{-\epsilon} = 1 - \epsilon lna 
+ O(\epsilon^{2})$, we can write
\begin{eqnarray}\label{A12}
\{\}^{-\epsilon}=  \{\}^{-\epsilon}|_{y=1} 
- \epsilon ln\Bigl[\frac{\{\}}{\{\}|_{y=1}}\Bigr].
\end{eqnarray}
Since the logarithm term vanishes when $y\rightarrow 1$, the remaining 
integral multiplied by this term is therefore convergent when 
$\epsilon \rightarrow 0$. The factor of $\epsilon$ multiplying the logarithm 
cancels the overall $\frac{1}{\epsilon}$ coefficient in $I_{4}(k_{i})$, 
contributes $O(\epsilon^{0})$ to that integral and shall be neglected 
henceforth. Utilizing this procedure, the three parametric integrals can be 
performed separately and expanding the results in $\epsilon$ leads to
\begin{eqnarray}\label{A13}
I_{4}(k_{i}) & = &
\frac{\mu^{-2\epsilon}}{2\epsilon^{2}}\left[1-\frac{1}{2}\epsilon-\epsilon 
\tilde{L}(P)\right].
\end{eqnarray}
\par Consider now the minimal number of diagrams belonging to the two-point 
vertex function, namely the integrals appearing in Eqs. (\ref{8}) 
and (\ref{9}). Denote the integral corresponding to Eq. (\ref{8}) (``sunset'') 
by $I_{3}(P)$, .i.e., the 
two-loop expression 
\begin{eqnarray}\label{A14}
I_{3}(P) & = & \int d^{d}q_{1}d^{d}q_{2} 
\frac{1}{(q_{1}^{2}+\mu^{2})(q_{2}^{2}+\mu^{2})[(q_{1}+q_{2}+P)^{2}+\mu^{2}]}.
\end{eqnarray}
Utilizing the partial $p$ technique defined by the operation 
(summation convention is implied and $i=1,...,d$, since the metric is 
Euclidean)
\begin{eqnarray}\label{A15}
1 = \frac{1}{2d}\;\left(\frac{\partial q_{1}^{i}}{\partial
q_{1}^{i}}+\frac{\partial q_{2}^{i}}{\partial q_{2}^{i}}\right)
\end{eqnarray}
we can rewrite last expression as
\begin{eqnarray}\label{A16}
I_{3}(P) & = &  \frac{1}{2d}\int
d^{d}q_{1}d^{d}q_{2}\left(\frac{\partial q_{1}^{\mu}}{\partial
q_{1}^{\mu}}+\frac{\partial q_{2}^{\mu}}{\partial q_{2}^{\mu}}\right)
\frac{1}{(q_{1}^{2}+\mu^{2})(q_{2}^{2}+\mu^{2})[(q_{1}+q_{2}+P)^{2}+\mu^{2}]}.
\end{eqnarray}
After integrations by parts and discarding surface terms we are led to 
\begin{eqnarray}\label{A17}
I_{3}(P)  & = &  -\frac{1}{d-3}[3\mu^{2}A(P)+B(P)],
\end{eqnarray}
where
\begin{subequations}
\begin{eqnarray}\label{A18}
A(P) & = & \int d^{d}q_{1}d^{d}q_{2}
\frac{1}{(q_{1}^{2}+\mu^{2})(q_{2}^{2}+\mu^{2})[(q_{1}+q_{2}+P)^{2}+\mu^{2}]^{2}},\label{A18a}\\
B(P) & = & \int d^{d}q_{1}d^{d}q_{2}
\frac{P . (q_{1}+q_{2}+P)}{(q_{1}^{2}+\mu^{2})(q_{2}^{2}+\mu^{2})[(q_{1}+q_{2}+P)^{2}+\mu^{2}]^{2}}\label{A18b}.
\end{eqnarray}
\end{subequations}
Let us first work out $A(P)$. We redefine the momenta in the following way: 
first we define a new momentum $-q_{1}^{'}=q_{1}+q_{2}$, such that 
$q_{1}= -(q_{1}^{'}+q_{2})$. Taking into account the invariance of the 
integral by the exchange $P \rightarrow -P$, after redefining back 
$q_{1}^{'}\rightarrow q_{1}$, $A(P)$ can be expressed in the form 
\begin{eqnarray}\label{A19}
A(P) & = & \int d^{d}q_{1}d^{d}q_{2}
\frac{1}{[(q_{1}+P)^{2}+\mu^{2}]^{2}(q_{2}^{2}+\mu^{2})[(q_{1}+q_{2})^{2}+\mu^{2}]}.
\end{eqnarray}
Introducing a Feynman parameter in order to solve the integral over $q_{2}$, 
using Eq. (\ref{A2}) and expanding everything in $d=4-\epsilon$ using the 
identity $\Gamma(a + b\epsilon)= \Gamma(a)\bigr[1+b\epsilon \psi(a) 
+ \frac{(b\epsilon)^{2}}{2}(\psi'(a) + \psi^{2}(a)) + O(\epsilon^{3})\bigl]$ 
(with $\psi(z)= \frac{dln\Gamma(z)}{dz}$), the last expression becomes  
\begin{eqnarray}\label{A20}
A(P)=\frac{1}{\epsilon}\left(1-\frac{1}{2}\epsilon + \frac{\pi^{2}}{24} 
\epsilon^{2}\right)\int_{0}^{1}dx[x(1-x)]^{-\epsilon/2}\int
\frac{d^{d}q_{1}}{[(q_{1}+P)^{2}+\mu^{2}]^{2}[q_{1}^{2}+m_{x}^{2}]^{\epsilon/2}}.
\end{eqnarray}
Employing another Feynman parameter, integrating over $q_{1}$ and expanding in 
$\epsilon$ as before, we have
\begin{eqnarray}\label{A21}
A(P) & = &
\frac{1}{4\epsilon}\left(1-\frac{1}{2}\epsilon + \frac{\pi^{2}}{24} 
\epsilon^{2}\right)\left(1-\frac{1}{2}\epsilon + \frac{\pi^{2}}{12} 
\epsilon^{2}\right) \int_{0}^{1}dx[x(1-x)]^{-\epsilon/2}\int_{0}^{1}
dy y^{\frac{\epsilon}{2}-1}\;\;\times \nonumber \\ & &(1-y)
\left\{y(1-y)P^{2}
+\left[1-y+\frac{y}{x(1-x)}\right]\mu^{2}\right\}^{-\epsilon}.
\end{eqnarray}
We wish to compute this integral up to its regular terms. Now, using the 
identity $y^{\frac{\epsilon}{2}-1}= 
\frac{2}{\epsilon}\frac{dy^{\frac{\epsilon}{2}}}{dy}$, integrating by parts 
over $y$ and keeping up to $O(\epsilon^{2})$ terms, we find
\begin{eqnarray}\label{A22}
A(P) & = &
\frac{\mu^{-2\epsilon}}{2\epsilon^{2}}\left(1-\frac{1}{2}\epsilon + 
\Bigr(\frac{\pi^{2}}{12}+\frac{1}{2}\Bigl) \epsilon^{2}\right) + \frac{\mu^{-2\epsilon}}{4} \tilde{i}(P),
\end{eqnarray}
where 
\begin{eqnarray}\label{A23}
&& \tilde{i}(P) = \int_{0}^{1} dx \int_{0}^{1}dy lny 
\frac{d}{dy}\Bigl((1-y)ln\Bigl[y(1-y)\frac{P^{2}}{\mu^{2}} + 1-y 
+ \frac{y}{x(1-x)} \Bigr]\Bigr).
\end{eqnarray}
\par The integral $B(P)$ can be rewritten as
\begin{eqnarray}\label{A25}
B(P) & = & -\frac{1}{2}P^{i}\frac{\partial}{\partial P^{i}}\int
d^{d}q_{1}d^{d}q_{2}
\frac{1}{(q_{1}^{2}+\mu^{2})(q_{2}^{2}+\mu^{2})[(q_{1}+q_{2}+P)^{2}+\mu^{2}]}.
\end{eqnarray}
Integrating over $q_{2}$ and performing a change of variables, we obtain
\begin{eqnarray}\label{A26}
B(P) & = &
-\frac{1}{2\epsilon}\left(1-\frac{1}{2}\epsilon\right)P^{i}\frac{\partial}{\partial
P^{i}}\int_{0}^{1}dx[x(1-x)]^{-\epsilon/2}\times \nonumber \\ & & \int
d^{d}q_{1}
\frac{1}{[(q_{1}+P)^{2}+\mu^{2}](q_{1}^{2}+m_{x}^{2})^{\epsilon/2}}.
\end{eqnarray}
Integration over $q_{1}$ followed by the expansion $d=4-\epsilon$ yields the 
result
\begin{eqnarray}\label{A27}
B(P) & = &
\frac{P^{2}}{4\epsilon}(1-\epsilon)\int_{0}^{1}dx[x(1-x)]^{-\epsilon/2}\int_{0}^{1}
dyy^{\epsilon/2}(1-y) \times \nonumber \\ & &
\left\{y(1-y)P^{2}+\left[1-y+\frac{y}{x(1-x)}\right]\mu^{2}\right\}^{-\epsilon}.
\end{eqnarray}
We expand the last bracket for small $\epsilon$ and when the parametric 
integrals are carried out, we find  
\begin{eqnarray}\label{A28}
B(P) & = & \frac{P^{2} \mu^{-2\epsilon}}{8\epsilon}\left[1-\frac{3}{4}\epsilon-2\epsilon
\tilde{L}_{3}(P,\mu)\right].
\end{eqnarray}
where
\begin{eqnarray}\label{A29}
\tilde{L}_{3}(P,\mu) & = &
\int_{0}^{1}dxdy(1-y)\ln\left\{y(1-y)\frac{P^{2}}{\mu^{2}} + 1-y 
+\frac{y}{x(1-x)}\right\}.
\end{eqnarray}
Consequently, the integral corresponding to the sunset diagram can be written 
as
\begin{eqnarray}\label{A30}
&& I_{3}(P)  =  \mu^{-2\epsilon}\Bigl \{
 -\frac{3\mu^{2}}{2\epsilon^{2}}\left[1+\frac{1}{2}\epsilon 
+ \Bigl(\frac{\pi^{2}}{12}+1\Bigr)\right] - \frac{3\mu^{2}}{4} \tilde{i}(P) \nonumber\\
&& \;\;\; - \frac{P^{2}}{8\epsilon}\Bigr[1+\frac{1}{4}\epsilon-2\epsilon
\tilde{L}_{3}(P,\mu)\Bigr]\Bigr\}.
\end{eqnarray}

This result implies that 
\begin{equation}\label{A31}
I_{3}(P)- I_{3}(P=0)  =  \mu^{-2\epsilon}\Bigl \{ - \frac{P^{2}}{8\epsilon}\Bigr[1+\frac{1}{4}\epsilon-2\epsilon
\tilde{L}_{3}(P,\mu)\Bigr] - \frac{3\mu^{2}}{4} \tilde{I}(P)\Bigr\},
\end{equation}
where

\begin{eqnarray}\label{A32}
&& \tilde{I}(P) = \int_{0}^{1} dx \int_{0}^{1}dy lny 
\frac{d}{dy}\Bigl((1-y)ln\Bigl[\frac{y(1-y)\frac{P^{2}}{\mu^{2}} + 1-y 
+ \frac{y}{x(1-x)}}{1-y 
+ \frac{y}{x(1-x)}} \Bigr]\Bigr).
\end{eqnarray}

This turns out to furnish the following diagrammatic expression:
\begin{eqnarray}\label{A33}
\left(
% sunset
\parbox{10mm}{\includegraphics[scale=1.0]{fig6.eps}}\bigg|_{\mu}  \quad - \quad
% sunset
\parbox{10mm}{\includegraphics[scale=1.0]{fig6.eps}}\bigg|_{k=0, \mu}\right) &=& \frac{N+2}{3} \mu^{-2\epsilon}\Bigl \{ - \frac{P^{2}}{8\epsilon}\Bigr[1+\frac{1}{4}\epsilon-2\epsilon
\tilde{L}_{3}(P,\mu)\Bigr]\nonumber\\ 
&& - \frac{3\mu^{2}}{4} \tilde{I}(P) \Bigr\}.
\end{eqnarray}
\par The integral $I_{5}(P)$ is defined by (see Eq. (\ref{9}))
\begin{eqnarray}\label{A34}
I_{5}(P) & = & \int
d^{d}q_{1}d^{d}q_{2}d^{d}q_{3}\frac{1}{(q_{1}^{2}+\mu^{2})(q_{2}^{2}+\mu^{2})(q_{3}^{2}+\mu^{2})}\times\nonumber
\\ & &
\frac{1}{[(q_{1}+q_{2}+P)^{2}+\mu^{2}][(q_{1}+q_{3}+P)^{2}+\mu^{2}]}.
\end{eqnarray}
The appropriate version of the partial $p$ procedure is now
\begin{eqnarray}\label{A35}
1 = \frac{1}{3d}\;\left(\frac{\partial q_{1}^{i}}{\partial
q_{1}^{i}}+\frac{\partial q_{2}^{i}}{\partial
q_{2}^{i}}+\frac{\partial q_{3}^{i}}{\partial q_{3}^{i}}\right).
\end{eqnarray}
Insert this identity inside the integrand, integrate by parts, 
get rid of surface terms and after some rearrangements we can write
\begin{eqnarray}\label{A36}
I_{5}(P)  & = &  -\frac{2}{3d-10}[5\mu^{2}C(P)+D(P)],
\end{eqnarray}
where
\begin{subequations}
\begin{eqnarray}\label{A37}
&& C(P) = \int d^{d}q_{1}d^{d}q_{2}d^{d}q_{3}
\frac{1}{(q_{1}^{2}+\mu^{2})(q_{2}^{2}+\mu^{2})(q_{3}^{2}+\mu^{2})}\times\nonumber
\\ 
&& \;\; \frac{1}{[(q_{1}+q_{2}+P)^{2}+\mu^{2}][(q_{1}+q_{3}+P)^{2}+\mu^{2}]^{2}}, \label{A37a}\\
&& D(P) = 
\left(-\frac{1}{2}P^{i}\frac{\partial}{\partial P^{i}}\right)\int
d^{d}q_{1}d^{d}q_{2}d^{d}q_{3}
\frac{1}{(q_{1}^{2}+\mu^{2})(q_{2}^{2}+\mu^{2})(q_{3}^{2}+\mu^{2})}\times\nonumber
\\ 
&&\;\;  \frac{1}{[(q_{1}+q_{2}+P)^{2}+\mu^{2}][(q_{1}+q_{3}+P)^{2}+\mu^{2}]}.\label{A37b}
\end{eqnarray}
\end{subequations}
Performing the replacement $q_{1}+P = q_{1}^{\prime}$, restoring 
$q_{1}^{\prime}\rightarrow q_{1}$ and $P\rightarrow -P$ just as we did 
before in the computation of $I_{3}(P)$, we have
\begin{subequations}
\begin{eqnarray}\label{A38}
C(P) & = & \int d^{d}q_{1}d^{d}q_{2}d^{d}q_{3}
\frac{1}{[(q_{1}+P)^{2}+\mu^{2}]^{2}}\times\nonumber \\ & &
\frac{1}{(q_{2}^{2}+\mu^{2})(q_{3}^{2}+\mu^{2})[(q_{1}+q_{2})^{2}+\mu^{2}][(q_{1}+q_{3})^{2}+\mu^{2}]},\label{A38a}\\
D(P) & = & -\frac{1}{2}P^{i}\frac{\partial}{\partial P^{i}}\int
d^{d}q_{1}d^{d}q_{2}d^{d}q_{3}
\frac{1}{[(q_{1}+P)^{2}+\mu^{2}]}\times\nonumber \\ & &
\frac{1}{(q_{2}^{2}+\mu^{2})(q_{3}^{2}+\mu^{2})[(q_{1}+q_{2})^{2}+\mu^{2}][(q_{1}+q_{3})^{2}+\mu^{2}]}.\label{A38b}
\end{eqnarray}
\end{subequations}
The object $C(P)$ can be rewritten in the form 
\begin{eqnarray}\label{A39}
C(P) & = & \int
d^{d}q_{1}\frac{1}{[(q_{1}+P)^{2}+\mu^{2}]^{2}}\times\nonumber \\ & &
\left(\int d^{d}q_{2}
\frac{1}{(q_{2}^{2}+\mu^{2})[(q_{1}+q_{2})^{2}+\mu^{2}]}\right)^{2}.
\end{eqnarray}
Now, following similar steps as those employed in the computation of $A(P)$, 
we get to the following result:
\begin{eqnarray}\label{A40}
C(P) & = &
\frac{\mu^{-3\epsilon}}{3\epsilon^{3}}\left(1-\frac{1}{2}\epsilon + 
\Bigr(\frac{\pi^{2}}{24}+\frac{9}{4}\Bigl) \epsilon^{2}\right) + \frac{\mu^{-3\epsilon}}{2\epsilon} \tilde{i}(P) ,
\end{eqnarray}
where $\tilde{i}(P)$ is given by Eq. (\ref{A23}). 
\par The integral $D(P)$ can be performed analogously and its singular 
part within the $\epsilon$ expansion reads:
\begin{eqnarray}\label{A41}
D(P) & = & \frac{P^{2}\mu^{-3\epsilon}}{6\epsilon^{2}}\left[1-\epsilon-3\epsilon
\tilde{L}_{3}(P,\mu)\right].
\end{eqnarray}
We can now express the  three-loop integral contributing to the two-point 
function in the form  
\begin{eqnarray}\label{A42}
I_{5}(P) & = &
 \mu^{-3\epsilon} \Bigl\{-\frac{5\mu^{2}}{3\epsilon^{3}}\Bigl[1+\epsilon 
+ \Bigl(\frac{\pi^{2}}{24} + \frac{15}{4} \Bigr) \epsilon^{2}\Bigr]
-\frac{5\mu^{2}}{2\epsilon} \tilde{i}(P) \nonumber\\
&& - \frac{P^{2}}{6\epsilon^{2}}\Bigl[1+\frac{1}{2}\epsilon-3\epsilon
\tilde{L}_{3}(P,\mu)\Bigr]\Bigr\}.
\end{eqnarray}
A useful quantity to our purposes is the difference between this integral 
computed at arbitrary external momentum from its value at $P=0$, namely
\begin{equation}\label{A43}
I_{5}(P)- I_{5}(P=0)  =  \mu^{-3\epsilon}\Bigl \{ - \frac{P^{2}}
{6\epsilon^{2}}\Bigr[1+\frac{1}{4}\epsilon-2\epsilon
\tilde{L}_{3}(P,\mu)\Bigr] - \frac{5\mu^{2}}{2\epsilon}
\tilde{I}(P) \Bigr\},
\end{equation} 
where the remaining integrals are the same as before, Eqs. (\ref{A32}). In 
summary, this results shall be useful in the definition of our unconventional
minimal subtraction in section III, but also in the standard BPHZ method 
using minimal subtraction. The latter requires, however, all diagrams from 
$\Gamma^{(2)}$, $\Gamma^{(4)}$. Further details can be found in the main text 
in connection with simplified versions of some results derived in 
this Appendix. 

\section{Computation of integrals useful in the BPHZ method}
\par  In this Appendix we shall calculate only three-loop diagrams which are 
momentum-dependent. The relevant graphs to be determined consist of the 
``sunset'' with a tadpole insertion and its counterterm, which is the sunset 
with a mass coupling constant generated iteratively from the BPHZ framework. As 
explicitly discussed in the body of the paper, all other three-loop 
``tadpole diagrams'' along with their counterterms do not depend on the 
external momenta. We shall not be concerned with their explicit calculation 
since they do not contribute to the field renormalization constant at 
three-loop order.  
\par Our aim will be modest here. We shall simply show that the singular part 
of the interesting diagram combined with its counterterm is 
momentum-independent. First, notice that the combination required of the 
two-point function of the diagram along with its tadpole is the following 
\begin{eqnarray}\label{B1} 
u^{3}\left[\frac{\mu^{3\epsilon}}{4} \parbox{12mm}{\includegraphics[scale=1.0]{fig8.eps}} \; + \; \frac{\mu^{2\epsilon} m^{2} \tilde{\lambda}_{m^ {2}}}{4u} \parbox{10mm}{\includegraphics[scale=1.0]{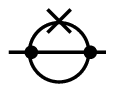}} \;\;\right].
\end{eqnarray}
The first diagram, the ``sunset with an inserted tadpole'', is given by the 
following expression 
\begin{eqnarray}\label{B2}
\parbox{10mm}{\includegraphics[scale=1.0]{fig8.eps}} = \Bigl[\frac{(N+2)}{3}\Bigr]^{2}i_{st} , 
\end{eqnarray}
whose associated integral reads 
\begin{eqnarray}\label{B3}
i_{st} &=& 
\int \frac{d^{d}q_{1}d^{d}q_{2}d^{d}q_{3}}{(q_{1}^{2}+m^{2})^{2}
(q_{2}^{2}+m^{2})((q_{1}+q_{2}+k)^{2}+m^{2})
(q_{3}^{2}+m^{2})}.
\end{eqnarray}
After integrating over $q_{3}$, it can be rewritten as
\begin{eqnarray}\label{B4}
i_{st} &=&-\frac{m^{2-\epsilon}}{\epsilon}
\int \frac{d^{d}q_{1}d^{d}q_{2}}{(q_{1}^{2}+m^{2})^{2}
(q_{2}^{2}+m^{2})((q_{1}+q_{2}+k)^{2}+m^{2})}.
\end{eqnarray}
Note that the remaining integral can be identified (after some reshuffling of 
the momenta) with $A(k)$ given by 
Eq. (\ref{A19}) when we replace $\mu \rightarrow m$. According to Eq. (\ref{A22}), we then 
obtain the following intermediate step for the diagram
\begin{eqnarray}\label{B5}
i_{st} &=&-\frac{m^{2-3\epsilon}}{2\epsilon^{3}}\left(1-\frac{1}{2}\epsilon + 
\Bigr(\frac{\pi^{2}}{12}+\frac{1}{2}\Bigl) \epsilon^{2}\right)  
- \frac{m^{2-3\epsilon}}{4\epsilon} \tilde{i}(k) .
\end{eqnarray}
Therefore, the total contribution of the first diagram is 
\begin{eqnarray}\label{B6}
\frac{\mu^{3\epsilon}}{4} \parbox{12mm}{\includegraphics[scale=1.0]{fig8.eps}} &=& - m^{2}\Bigl(\frac{m}{\mu}\Bigr)^{-3\epsilon} 
\frac{(N+2)^{2}}{72}\Bigl[\frac{1}{\epsilon^{3}}\Bigl(1-\frac{1}{2}\epsilon + 
\Bigr(\frac{\pi^{2}}{12}+\frac{1}{2}\Bigl) \epsilon^{2}\Bigr) \nonumber \\
&& \;\; + \;\; \frac{1}{2\epsilon} \tilde{i}(k) \Bigr].
\end{eqnarray}
The counterterm diagram turns out to be written as 
\begin{eqnarray}\label{B7}
\parbox{12mm}{\includegraphics[scale=1.0]{fig27.eps}} = \frac{(N+2)}{3}A(k).
\end{eqnarray}
After replacing the value $\tilde{\lambda}_{m^{2}}=u\delta_{m^{2}}^{(1)}= 
\frac{(N+2)u}{6\epsilon}$ and using the previous expression for $A(k)$ 
Eq. (\ref{A22}) , the overall contribution of the counterterm is easy to determine, 
namely
\begin{eqnarray}\label{B8}
\frac{\mu^{2\epsilon} m^{2} \tilde{\lambda}_{m^ {2}}}{4u} \parbox{11mm}{\includegraphics[scale=1.0]{fig27.eps}} &=& m^{2}\Bigl(\frac{m}{\mu}\Bigr)^{-2\epsilon} 
\frac{(N+2)^{2}}{72}\Bigl[\frac{1}{\epsilon^{3}}\Bigl(1-\frac{1}{2}\epsilon + 
\Bigr(\frac{\pi^{2}}{12}+\frac{1}{2}\Bigl) \epsilon^{2}\Bigr) \nonumber \\
&& \;\; + \;\; \frac{1}{2\epsilon} \tilde{i}(k) \Bigr].
\end{eqnarray}
Therefore, summing up (\ref{B6}) and (\ref{B8}), the singular terms which depend on the 
external momenta exactly cancel each other. Indeed, performing explicitly the 
summation, the aforementioned combination of these two diagrams yields
\begin{eqnarray} \label{B9}
u^{3}\left[\frac{\mu^{3\epsilon}}{4} \parbox{10mm}{\includegraphics[scale=1.0]{fig8.eps}} \; + \; \frac{\mu^{2\epsilon} m^{2} \tilde{\lambda}_{m^ {2}}}{4u} \parbox{9mm}{\includegraphics[scale=1.0]{fig27.eps}} \;\;\right] &=& m^{2}ln\Bigl(\frac{m}{\mu}\Bigr) 
\frac{(N+2)^{2}}{72}\Bigl[\frac{1}{\epsilon^{2}}\Bigl(1-\frac{1}{2}\epsilon + 
\Bigl(\frac{\pi^{2}}{12}+\frac{1}{2}\Bigr) \epsilon^{2}\Bigr)\nonumber\\
&& \qquad \qquad \times \Bigl\{1-\frac{5\epsilon}{2}ln\Bigl(\frac{m}{\mu}\Bigr)\Bigl\}\Bigr]. 
\end{eqnarray}
\par Although there are singular terms 
proportional to $ln\Bigl[\frac{m}{\mu}\Bigr]$ as explained before, 
the contributions coming from all three-loop diagrams of the two-point 
vertex part shall eliminate them, precisely as we have shown 
explicitly at two-loop level, although we do not pursue this proof herein. 
Since our goal is just to collect singular terms which are explicitly 
momentum-dependent for the reasons explained in the main text, the 
terms multiplying $ln\Bigl[\frac{m}{\mu}\Bigr]$ can 
be safely neglected in the computation of the field normalization function 
at three-loop order. This concludes our task.

\end{document}